\documentclass[preprint,12pt]{elsarticle}

\journal{Computer Networks}

\usepackage{amssymb}
\usepackage{amsmath}
\usepackage{bigints}

\usepackage{array}
\usepackage{booktabs}
\usepackage{multirow}
\usepackage{subcaption}
\usepackage{float}

\usepackage{hyperref}
\usepackage{url}

\begin{document}

\begin{frontmatter}
\title{Temporal Analysis Framework for Intrusion Detection Systems: A Novel Taxonomy for Time-Aware Cybersecurity}


\author[inst1,inst2]{Tatiana S. Parlanti\corref{cor1}}
\ead{tatiana.parlanti@ingenieria.uncuyo.edu.ar}

\author[inst1,inst2]{Carlos A. Catania}
\ead{harpo@ingenieria.uncuyo.edu.ar}

\cortext[cor1]{Corresponding author.}

\affiliation[inst1]{organization={Universidad Nacional de Cuyo, Facultad de Ingenier{\'i}a, Laboratorio de Sistemas Inteligentes (LABSIN)},
            addressline={Centro Universitario},
            city={Mendoza},
            postcode={M5502JMA},
            state={Mendoza},
            country={Argentina}}

\affiliation[inst2]{organization={Consejo Nacional de Investigaciones Cient{\'i}ficas y T{\'e}cnicas (CONICET)},
            addressline={Godoy Cruz 2290},
            city={C.A.B.A},
            postcode={C1425FQB},
            state={Buenos Aires},
            country={Argentina}}

\tnotetext[1]{Preprint submitted to Computer Networks}

\begin{abstract}
Most intrusion detection systems still identify attacks only after significant damage has occurred, detecting late-stage tactics rather than early indicators of compromise. This paper introduces a temporal analysis framework and taxonomy for time-aware network intrusion detection. Through a systematic review of over 40 studies published between 2020 and 2025, we classify NIDS methods according to their treatment of time, from static per-flow analysis to multi-window sequential modeling. The proposed taxonomy reveals that inter-flow sequential and temporal window-based  methods provide the broadest temporal coverage across MITRE ATT\&CK tactics, enabling detection from Reconnaissance through Impact stages. Our analysis further exposes systematic bias in widely used datasets, which emphasize late-stage attacks and thus limit progress toward early detection. This framework provides essential groundwork for developing IDS capable of anticipating rather than merely reacting to cyber threats, advancing the field toward truly proactive defense mechanisms.
\end{abstract}




\end{frontmatter}


\section{Introduction}

State-of-the-art Network Intrusion Detection Systems described in the literature face a temporal paradox: they excel at identifying attacks already \textit{in progress}, i.e., when data is being exfiltrated, lateral movement concludes, or ransomware executes, corresponding to the \textit{Impact} and other late-stage tactics in the MITRE ATT\&CK framework~\cite{mitreattack}. Yet, they remain far less effective at capturing the subtle temporal patterns that characterize earlier adversarial behaviors, when detection would enable true prevention. This imbalance does not necessarily reflect poor model quality, but rather how intrusion detection has been conceptualized and assessed: systems are optimized to recognize clear, high-impact manifestations of compromise, rather than the gradual temporal evolution that precedes them.

As noted by previous studies~\cite{kimAIIDSApplicationDeep2020,wang_robust_2023,o_lopes_network_2023}, incorporating temporal information allows the detection of attacks that evolve over time or follow specific temporal distributions. Capturing the order and timing of events can reveal anomalies invisible in static analysis, since network traffic inherently forms a sequence of time-stamped interactions. Furthermore, by capturing how network behaviors evolve over time, temporal modeling also provides a foundation for assessing dynamic trust relationships among network entities. However, the field still lacks a coherent framework that explains how different temporal modeling strategies affect detection across the adversarial lifecycle. The central question, therefore, is whether explicitly modeling temporality can shift detection capabilities toward earlier attack stages.

To address this question, we conduct an extensive survey and propose a temporal taxonomy that categorizes NIDS approaches according to their treatment of time—ranging from static per-flow models to sequential, window-based analysis of evolving network states. Using this taxonomy as an analytical framework, we examine over 40 recent publications (2020–2025) to assess current capabilities.

Our analysis yields three key insights. First, the temporal modeling capabilities vary across existing approaches in the literature. However, to the best of our knowledge, there has been no unified framework to compare or classify them. As a result, there is no conclusive way to assess the current state of temporal detection in NIDS. The proposed taxonomy addresses this gap by defining five categories: Static Per-Flow, Static Contextual Snapshot, Intra-Flow Sequential, Inter-Flow Sequential, and Temporal Window-Based Sequential Analysis. Second, different temporal strategies align with specific stages of the MITRE ATT\&CK lifecycle, offering distinct potential for intrusion detection. Third, widely used benchmark datasets tend to emphasize late-stage behaviors, constraining progress toward proactive intrusion detection.

The major contributions of the paper are as follows.
\begin{description}
    \item[Novel Temporal Taxonomy.] A five-category classification (S.1–T.3) that organizes time-aware NIDS approaches by their temporal analysis capabilities.
    \item[ATT\&CK Alignment.] An analytical mapping linking the proposed taxonomy to the MITRE ATT\&CK framework, clarifying the coverage of temporal approaches across more than 40 attack techniques.
    \item[Dataset Bias Recognition.] Identification of inherent biases in benchmark datasets (CIC-IDS2017, UNSW-NB15, NSL-KDD) that skew evaluation toward Impact-phase attacks.
    \item[Systematic Review.] A detailed analysis of recent work (2020–2025), highlighting strengths, limitations, and research directions for time-aware intrusion detection. 
\end{description}

The rest of this document is organized as follows. Section~\ref{sec:search-criteria} exhibits the methodology used for the selection criteria of the papers presented in
this survey. Section~\ref{sec:background} presents fundamental concepts, then in Section~\ref{sec:related-works} the related works are presented. In Section~\ref{sec:taxonomy} a taxonomy is proposed for analyzing the literature, which is effectively carried out in Section~\ref{sec:literature-review}. The discussions and main limitations are then presented in Sections~\ref{sec:discussion} and~\ref{sec:open-issues}, respectively. Finally, the conclusions are presented in Section~\ref{sec:conclusions}.

\section{Methodology} \label{sec:search-criteria}

This systematic literature review examines the application of temporal and dynamic approaches in network intrusion detection systems, covering publications from 2020 through October 2025. The review process was designed to identify the most relevant studies while ensuring balanced coverage across the field.

\subsection{Search Strategy and Information Resources}

The literature search was conducted across multiple complementary academic databases to capture diverse perspectives and publication venues in the field. The initial exploration was performed using the Crossref API~\cite{CrossrefAPI}, chosen for its broad coverage across publishers and venues. The search was then expanded to include EBSCO~\cite{EBSCOhost} for additional journal coverage, Scopus~\cite{Scopus} for its comprehensive indexing and citation analytics, IEEE Xplore~\cite{IEEEXplore} for publications in computer science and engineering, and the ACM Digital Library~\cite{ACMDL} for computing-focused research. 

\subsection{Query Construction and Filtering Process}

Search terms were organized around three conceptual pillars. First, terminology related to temporal characteristics, such as \textit{temporality, temporal, time-sensitive, time-series, dynamic}. Second, terms that focus on the domain scope, like \textit{intrusion detection systems, IDS, NIDS}. Finally, methodological approaches such as recurrent and sequential architectures, and graph-based methods were included. This conceptual structure allowed for capturing both traditional and emerging temporal approaches without restricting the search to a single technical paradigm.

To ensure research quality and relevance, the selection was limited to peer-reviewed journal articles and conference proceedings published in Q1 or Q2 venues according to the Scimago Journal Rankings~\cite{ScimagoJR}. Each paper was required to explicitly address intrusion detection or network security and incorporate a temporal, dynamic, or sequential analytical component. We excluded works focused exclusively on domain-specific applications such as vehicular networks, medical monitoring, industrial control systems, or video surveillance, as well as preliminary or duplicate publications.

\subsection{Selection and Analysis Process}

The selection process followed an iterative refinement cycle, evolving as insights emerged about how temporality is incorporated into intrusion detection research. Initial automated searches, particularly those via Crossref, yielded a broad pool of candidate studies; however manual inspection revealed that many addressed temporal aspects only marginally. Recognizing this, subsequent searches in EBSCO, Scopus, IEEE Xplore, and ACM Digital Library were refined and the methodology sections were manually examined to emphasize studies where temporal modeling played a central role.

This iterative process, combining automated retrieval, conceptual refinement, and manual verification, resulted in a final corpus of 44 publications. The corpus includes both classical methods and recent innovations, providing a comprehensive foundation for understanding the current state and trajectory of the field.

\section{Background} \label{sec:background}
An intrusion is any unauthorized activity that occurs within a network or system, which may or may not be malicious in intent~\cite{rabashNonDominatedSortingGenetic2023}. When such activity aims to cause harm, such as stealing data, corrupting files, or spreading malware, it becomes a cyberattack. To identify and mitigate these events, Intrusion Detection Systems (IDSs) continuously monitor network or host activity for signs of abnormal or unauthorized behavior~\cite{al-ghuwairiIntrusionDetectionCloud2023}. Among them, Network-based IDSs (NIDSs) analyze network traffic to detect threats and malicious communication patterns~\cite{fengWeightedIntrusionDetection2021}.

Traditional NIDS approaches, such as signature-based or rule-based detection, rely on predefined patterns of known attacks~\cite{dengEdgefeaturedMultihopAttention2025,xuApplyingSelfsupervisedLearning2024,halbouniCNNLSTMHybridDeep2022}. While effective for well-understood threats, these methods fail to recognize novel or evolving attack strategies and require frequent manual updates~\cite{cai_malicious_2024,fengWeightedIntrusionDetection2021}. As modern networks face increasingly complex and adaptive threats, machine learning (ML)-based NIDS have emerged as a powerful alternative~\cite{dengEdgefeaturedMultihopAttention2025,psychogyiosDeepLearningIntrusion2024,sethNovelTimeEfficient2021}. These systems can learn intricate statistical or behavioral patterns directly from data, enabling them to generalize to unseen attacks and reduce the dependence on handcrafted signatures.

Despite persistent challenges~\cite{saikamEESNNHybridDeep2024,ren_canet_2023,cai_malicious_2024}, such as false positives, data quality, feature selection, interpretability, and computational overhead, ML-based NIDS remain among the most promising approaches for modern intrusion detection. Accordingly, this survey focuses on machine learning-based Network Intrusion Detection Systems, emphasizing recent trends toward temporal and sequential analysis.

In ML-based NIDS, the effectiveness of the model heavily depends on data preprocessing and feature selection~\cite{sethNovelTimeEfficient2021}. Since network traffic data are typically high-dimensional and heterogeneous, selecting relevant features is essential to reduce noise, avoid overfitting, and improve detection accuracy. Common feature selection strategies include filter~\cite{halbouniCNNLSTMHybridDeep2022,emanetEnsembleLearningBased2023,psychogyiosDeepLearningIntrusion2024}, wrapper~\cite{wanjau_discriminative_2023,sethNovelTimeEfficient2021,emanetEnsembleLearningBased2023}, and embedded~\cite{saikamEESNNHybridDeep2024} methods, which differ in how they evaluate the contribution of each feature to the model’s performance. 

In addition to feature selection, feature engineering often involves transforming or constructing new variables that better represent underlying traffic behaviors, enabling learning algorithms to capture discriminative patterns more effectively. Likewise, dimensionality reduction techniques such as PCA~\cite{niknamiEnhancedMetaIDSAdaptive2025,wanjau_discriminative_2023}, LDA~\cite{han_network_2023}, t-SNE~\cite{da_silva_ruffo_generative_2025}, and UMAP~\cite{lo2021egraphsage} are commonly applied to simplify data representations, decrease computational cost, and enhance model generalization. These steps, while not the focus of this survey, constitute the foundation upon which many temporal and sequential detection models are built.

\subsection{Learning Approaches}

A wide range of machine learning (ML) techniques has been employed to model and classify network traffic in NIDS. Classical ML methods have long been used to detect anomalies and classify traffic patterns based on handcrafted features. More recently, deep learning (DL) techniques have enabled the automatic extraction of complex hierarchical representations from raw or minimally processed data. Also, graph-based deep learning models have emerged as a promising paradigm for capturing the structural and relational dependencies inherent in network traffic, representing connections among hosts, flows, or packets as graphs that reflect the topology and dynamics of the network.

\paragraph{Machine Learning (ML)} Various traditional classification algorithms have been used to build NIDS, such as: 
\begin{description}
    \item Logistic Regression~\cite{niknamiEnhancedMetaIDSAdaptive2025,emanetEnsembleLearningBased2023}: Linear model for binary classification that estimates the probability of a class using the logistic (sigmoid) function on a linear combination of features.
    \item Decision Tree (DT)~\cite{emanetEnsembleLearningBased2023}: Supervised learning model in which data is divided according to its features, forming a tree-like structure for decision making, where the leaves represent each of the classes and the branches represent conjunctions of features that lead to those labels.
    \item Random Forest (RF)~\cite{sethNovelTimeEfficient2021}: An ensemble of multiple decision trees built on random subsets of data or features, combining the predictions of each tree to produce the final decision.
    \item Gradient Boosting (GB)~\cite{sethNovelTimeEfficient2021}: An ensemble technique where models are built sequentially, with each new model correcting the errors of the previous one. In practice, decision trees are often used as base learners, but other models can also be employed. There are several GB-based algorithms, such as LightGBM~\cite{sethNovelTimeEfficient2021,jinSwiftIDSRealtimeIntrusion2020}, which is optimized to handle large volumes of high-dimensional data.
    \item $k-$Nearest Neighbors (KNN)~\cite{sethNovelTimeEfficient2021}: A classifier that assigns a class to a new observation based on the classes of its $k$ nearest neighbors in the feature space.
    \item Genetic Algorithms (GA)~\cite{catania2013towards,zhangModelIntrusionDetection2020}: An optimization method inspired by biological evolution, which represents possible solutions using encodings (chromosomes) and applies evolutionary operators such as natural selection, crossover, and/or mutation to iteratively evolve toward optimal or near-optimal solutions to a problem.
    \item Support Vector Machine (SVM)~\cite{fengWeightedIntrusionDetection2021}: A classification algorithm that finds the hyperplane that best separates the classes, maximizing the margin between the closest data points of each class, which are called support vectors.
    \item Multi-Layer Perceptron (MLP)~\cite{emanetEnsembleLearningBased2023}: A type of artificial neural network composed of multiple layers of neurons with nonlinear activation functions, capable of learning complex representations of data.
\end{description}

\paragraph{Deep Learning (DL)} This is a subfield of ML that uses neural networks with multiple layers to learn complex data representations. Therefore, unlike previous algorithms that rely on feature engineering, DL models can automatically extract meaningful representations from data. Different neural network architectures have been applied in the context of NIDSs, such as:
\begin{description}
    \item Deep Neural Network (DNN)~\cite{han_network_2023,thirimanneDeepNeuralNetwork2022}: An artificial neural network with multiple hidden layers between the input and output.
    \item Convolutional Neural Network (CNN)~\cite{halbouniCNNLSTMHybridDeep2022,kimAIIDSApplicationDeep2020}: A type of neural network designed to extract spatial features from data, using filters (convolutions) that detect local patterns. These networks are especially useful for data with an Euclidean structure, such as images, where the spatial proximity between pixels is relevant.
    \item Capsule Networks (CapsNets)~\cite{logeswari2025comprehensive}: use capsules, i.e., groups of neurons that work together to represent specific features or patterns instead of individual neurons. 
    \item Recurrent Neural Network (RNN)~\cite{zavrakFlowbasedIntrusionDetection2023}: A neural network designed to process sequential data, such as text or time series, as it has an internal ``memory'' that allows it to use information from previous steps when making a prediction. Some widely used RNNs are Long Short-Term Memory (LSTM)~\cite{shang_discovering_2021,kimAIIDSApplicationDeep2020}, designed to remember long-term information in data sequences; Bidirectional LSTM (BiLSTM)~\cite{wu_active_2024,wanjau_discriminative_2023}, which is a variant that processes a sequence in both directions: forward and backward, so it cannot be used in real time; and Gated Recurrent Unit (GRU)~\cite{duan2023application,king2023euler}, which uses gates to control what information to remember or forget in a sequence.
    \item Temporal Convolutional Network (TCN)~\cite{da_silva_ruffo_generative_2025}: A convolutional neural network designed to model sequences over time, using causal 1D convolutions, which ensure that the output at time $t$ only depends on inputs up to that point, and not on the future. In contrast, Bidirectional TCN (BiTCN)~\cite{cai_malicious_2024,chen_novel_2023} is a variant that processes the sequence in both directions: from past to future, and from future to past, so it cannot be used in real time. Unlike RNNs, TCN and BiTCN are fully convolutional architectures, without recurrence.      
    \item Generative Adversarial Network (GAN)~\cite{da_silva_ruffo_generative_2025,saikamEESNNHybridDeep2024}: A model composed of two neural networks, called generator and discriminator, which compete with each other, as the former generates false data and the latter tries to distinguish it from the real data, both improving over time.
\end{description}

\noindent In addition, some approaches incorporate attention mechanisms, which is a technique that allows a network to focus on the most relevant parts of the input when processing it, rather than treating all information equally.

\paragraph{Graph-Based Deep Learning}Taking advantage of the topological relationships in the network, in recent years, graph neural networks (GNNs) have been used for intrusion detection~\cite{zhong2024survey,bilot2023graph}. A graph is a set of nodes connected by edges, representing entities and their relationships. Unlike Euclidean data, which is structured in regular grids or coordinate spaces (as in images or time series), graph data captures irregular and non-Euclidean relationships, where connections depend on the network topology rather than fixed spatial proximity. In the context of NIDS, a graph consists of nodes (e.g., IPs or flows) and edges representing communication or dependency relationships. 

\noindent A GNN is a neural network designed to work with data that has a graph structure. It learns representations of nodes or edges by aggregating information from each node's neighbors, thereby capturing both the structural context and feature attributes of the graph. Among the most widely used GNN architectures are, for example: 

\begin{description}
    \item Graph Convolutional Network (GCN)~\cite{kipf2016semi}: A type of GNN that performs a convolution-like operation on graphs, aggregating information from each node and its neighbors in a weighted manner. 
    \item Graph Convolutional Network II (GCNII)~\cite{duan2023application}: An improved version of GCN that avoids information oversaturation when many layers are added. It introduces residuals and initial feature preservation to allow for deeper networks.
    \item Graph Sample and Aggregate (GraphSAGE)~\cite{hamilton2017graphSAGE}: A GNN model that takes a sample of the set of neighbors for each node and then aggregates them to update their representations.
    \item Edge-Enhanced GraphSAGE (E-GraphSAGE)~\cite{lo2021egraphsage}: An extension of GraphSAGE that incorporates information from edges in addition to nodes.
    \item Temporal GNN (Temporal Graph Neural Network)~\cite{duan2023application,king2023euler,rossi2020temporal}: A type of neural network designed to process graph data that changes over time. Unlike traditional GNNs, which work on static structures, temporal GNNs model the dynamic evolution of nodes, edges, or attributes of the graph in a temporal sequence.
\end{description}

It is worth noting that among the learning approaches, the following stand out: \textit{supervised learning}~\cite{dengEdgefeaturedMultihopAttention2025,wanjau_discriminative_2023}, in which models learn from previously labeled data; \textit{semi-supervised learning}~\cite{da_silva_ruffo_generative_2025,zavrakFlowbasedIntrusionDetection2023}, in which there is a mixture of labeled and unlabeled data; \textit{self-supervised learning}~\cite{xuApplyingSelfsupervisedLearning2024}, where the model generates its own labels from the data; \textit{active learning}~\cite{wu_active_2024}, in which the model selects which data should be labeled, optimizing the training process; and \textit{unsupervised learning}~\cite{wang_robust_2023}, where the data is unlabeled and the model finds patterns on its own.

\subsection{Common Metrics}

The performance of ML-based NIDS is commonly evaluated using a set of quantitative metrics that depend on the detection objective and data characteristics. When true class labels are available, a confusion matrix  can be constructed to summarize the relationship between predicted and actual outcomes (see Table~\ref{tab:confusion-matrix}). In binary classification, it records the number of true positives (TP), false negatives (FN), true negatives (TN), and false positives (FP), which form the basis for most evaluation metrics. Based on the confusion matrix for binary classification, Table \ref{tab:metrics-clasif} defines the main metrics.

\begin{table}[h]
    \caption{Confusion matrix for binary classification.}
    \label{tab:confusion-matrix}
    \centering
    \scriptsize
    \renewcommand{\arraystretch}{1.0}
    \begin{tabular}{@{}c|cc@{}}
        \toprule
         & \textbf{Predicted Positive} & \textbf{Predicted Negative} \\
         \midrule
        \textbf{Actual Positive} & TP & FN \\
        \textbf{Actual Negative} & FP & TN \\
        \bottomrule
    \end{tabular}
\end{table}

\begin{table}
    \caption{Metrics for evaluating classification models.}
    \label{tab:metrics-clasif}
    \centering
    \scriptsize
    \begin{tabular}{
      @{}
      >{\centering\arraybackslash}p{2.5cm} 
      c
      >{\raggedright\arraybackslash}p{5.5cm} 
      >{\centering\arraybackslash}p{1.5cm}
      @{}
    }
    \toprule
    \textbf{Metric} & \textbf{Formula} & \textbf{Description} & \textbf{Example of use} \\
    \midrule
    \textbf{Accuracy} & $\phantom{\bigintsss\!} \frac{TP + TN}{TP + TN + FP + FN}$ & Total correct predictions of the model. & \cite{niknamiEnhancedMetaIDSAdaptive2025} \\
    \textbf{Precision} & $\phantom{\bigintsss\!} \frac{TP}{TP + FP}$ & Of the alerts reported, how many corresponded to actual intrusions. & \cite{da_silva_ruffo_generative_2025} \\
    \textbf{Recall (True Positive Rate, Detection Rate, Sensitivity)} 
    & $\phantom{\bigintsss\!} \frac{TP}{TP + FN}$ & Of the actual number of intrusions, how many were correctly identified (correct alerts). & \cite{da_silva_ruffo_generative_2025} \\ 
    \textbf{F1-Score} & $\phantom{\bigintsss\!} 2 \cdot \frac{\text{Precision} \cdot \text{Recall}}{\text{Precision} + \text{Recall}}$ & Balance between precision and recall. Useful when there are unbalanced classes. & \cite{da_silva_ruffo_generative_2025} \\
    \textbf{False Positive Rate (False Alarm Rate)} & $\phantom{\bigintsss\!} \frac{FP}{FP + TN}$ & Of the total number of benign events, how many were mistakenly classified as intrusions. & \cite{saikamEESNNHybridDeep2024} \\ 
    \textbf{False Negative Rate} & $\phantom{\bigintsss\!} \frac{FN}{FN + TP}$ & Of the actual number of intrusions, how many were not detected (no alert was issued). & \cite{rajeshkannaUnifiedDeepLearning2021} \\ 
    \textbf{Specificity (True Negative Rate)} & $\phantom{\bigintsss\!} \frac{TN}{TN + FP}$ & Of the total number of benign events, how many were correctly identified (not confused with intrusions). & \cite{sethNovelTimeEfficient2021} \\
    \textbf{Area Under ROC Curve (AUC-ROC)} & $\bigintsss\! ROC$ & Area under the TPR vs. FPR curve (ROC curve) obtained by varying the decision threshold between 0 and 1. & \cite{dengEdgefeaturedMultihopAttention2025} \\ 
    \textbf{Area Under PR Curve (AUC-PR)} & $\bigintsss\! PR$ & Area under the Precision vs. Recall curve (PR curve) obtained by varying the decision threshold between 0 and 1. & - \\ 
    \bottomrule
    \end{tabular}
\end{table}

For multi-class scenarios, an $n \times n$ confusion matrix is used, where $n$ is the number of classes. Here, metrics can be averaged using macro, micro, or weighted aggregation strategies~\cite{wang_spatial-temporal_2024,dengEdgefeaturedMultihopAttention2025,xuApplyingSelfsupervisedLearning2024}. Macro-averaging computes each class metric independently and then averages them, while micro-averaging aggregates all TP, FP, FN, and TN before calculation; the weighted variant adjusts each class’s contribution based on class frequency.

In addition, the AUC-ROC and AUC-PR represent the area under the ROC and Precision-Recall (PR) curves, respectively~\cite{davis2006-auc-roc-pr}. The ROC curve shows the tradeoff between the true positive rate (TPR) and false positive rate (FPR), while the PR curve illustrates the tradeoff between Precision and Recall, as the decision threshold varies. AUC-ROC reflects the model's overall ability to distinguish between classes, and can be interpreted as the probability that a randomly chosen positive instance receives a higher score than a randomly chosen negative one. In contrast, AUC-PR evaluates the reliability of positive predictions, which is valuable for imbalanced classification problems. However, despite the fact that intrusion detection is inherently imbalanced, AUC-PR has not been observed in the literature review conducted in this work, unlike AUC-ROC~\cite{dengEdgefeaturedMultihopAttention2025,psychogyiosDeepLearningIntrusion2024}.

Finally, especially in real-time detection problems, computing time~\cite{emanetEnsembleLearningBased2023}, training and prediction time~\cite{sethNovelTimeEfficient2021}, CPU time~\cite{niknamiEnhancedMetaIDSAdaptive2025}, and memory consumption~\cite{han_network_2023} are also reported.

\subsection{Adversarial Tactics} \label{sec:mitre}

The performance metrics and learning approaches discussed above ultimately aim to improve a system’s ability to detect and characterize adversarial behavior within network environments. To provide a common and systematic way to describe such adversarial behaviors, the MITRE ATT\&CK (Adversarial Tactics, Techniques, and Common Knowledge) framework~\cite{mitreattack} offers a comprehensive taxonomy of real-world attack tactics and techniques. MITRE ATT\&CK organizes adversarial actions into tactics, which represent the attacker’s high-level objectives, and techniques, which describe the specific methods used to achieve each goal. 

Table~\ref{tab:mitre} describes each tactic defined by MITRE ATT\&CK for the Enterprise domain, which covers behavior against enterprise IT networks and cloud. It is ordered according to the typical progression of an intrusion, from planning to execution and exploitation. This framework provides a conceptual foundation for evaluating how well NIDS models can detect attacks across different stages of the adversarial lifecycle.

\begin{table}[h]
    \caption{ATT\&CK Matrix for Enterprise.}  
    \label{tab:mitre}
    \centering
    \scriptsize
    \begin{tabular}{@{}
    >{\raggedleft\arraybackslash}p{3cm} 
      >{\raggedright\arraybackslash}p{10cm}
      @{}} 
    \toprule    
    \textbf{Name} & \textbf{Description} \\
    \midrule
    Reconnaissance & The adversary is trying to gather information they can use to plan future operations. \\
    Resource Development & The adversary is trying to establish resources they can use to support operations. \\
    Initial Access & The adversary is trying to get into the victim's network. \\
    Execution & The adversary is trying to run malicious code. \\
    Persistence & The adversary is trying to maintain their foothold. \\
    Privilege Escalation & The adversary is trying to gain higher-level permissions. \\
    Defense Evasion & The adversary is trying to avoid being detected. \\
    Credential Access & The adversary is trying to steal account names and passwords. \\
    Discovery & The adversary is trying to figure out the victim's environment. \\
    Lateral Movement & The adversary is trying to move through the victim's environment. \\
    Collection & The adversary is trying to gather data of interest to their goal. \\
    Command and Control (C\&C or C2) & The adversary is trying to communicate with compromised systems to control them. \\
    Exfiltration & The adversary is trying to steal data. \\
    Impact & The adversary is trying to manipulate, interrupt, or destroy the victim's systems and data.  \\
    \bottomrule
    \end{tabular}
    \raggedright
    \textit{Note}: Table adapted from \cite{mitreattack}.
\end{table}

\section{Related Works} \label{sec:related-works}

The field of Network Intrusion Detection Systems (NIDS) has been extensively studied, leading to numerous comprehensive surveys. Some of these works address deep learning applications in cybersecurity from a broader perspective, such as \cite{macasSurveyDeepLearning2022a}, which includes intrusion detection, among others. However, many other works focus specifically on how Machine Learning (ML) and Deep Learning (DL) algorithms have been applied to intrusion detection. For instance, \cite{habeeb2022network} analyze AI-based NIDS solutions proposed between 2016 and 2021, evaluating their strengths, limitations, adopted AI methodologies, datasets, and performance metrics. Similarly, \cite{abdulganiyu2023systematic} review anomaly-, signature-, and hybrid-based NIDS from 2014 to 2022, emphasizing the growing trend toward deep learning and anomaly-based approaches. In \cite{kimanziDeepLearningAlgorithms2024}, the authors investigate the use of deep learning techniques in IDS, such as Convolutional Neural Networks (CNN), Recurrent Neural Networks (RNN), Deep Neural Networks (DNN), and Long Short-Term Memory (LSTM), among others, covering studies from 2019 to 2023. Likewise, \cite{maseerMetaanalysisSystematicReview2024} examine the application of ML and DL methods in anomaly-based NIDS between 2017 and 2022. Recent reviews have expanded this perspective, for example \cite{zhangReviewDeepLearning2025} synthesize the use of deep learning in IDS with emphasis on spatiotemporal feature extraction and data imbalance. Furthermore, in \cite{zhong2024survey}, the authors analyze GNN-based intrusion detection models. Complementary studies like \cite{goldschmidt2025network} analyze the reliability and composition of public NIDS datasets, identifying recurring issues and exploring long-term research and development directions. Finally, to contextualize threat defense strategy, \cite{jiangMITREATTCKApplications2025} provide a systematic review on the extensive application of the MITRE ATT\&CK framework across various cybersecurity domains, including intrusion detection, threat intelligence, and vulnerability analysis.

While these works have been instrumental in organizing the methodological landscape, they share a common limitation: the temporal behavior of network traffic is rarely treated as a central analytical dimension. Although \cite{zhangReviewDeepLearning2025} acknowledges temporal dependencies through its focus on spatiotemporal feature extraction, its approach remains primarily technical, centered on deep learning architectures and feature modeling. In contrast, our study treats temporality as a structural property of intrusion detection, proposing a taxonomy that classifies systems according to the way they encode temporal information, independent of the underlying learning paradigm. This conceptual perspective aligns with emerging research that recognizes time as a key signal in cyber defense. For instance, \cite{landauer_review_2025} study time-series analytics in cybersecurity analytics, classifying temporal characteristics (e.g., granularity and seasonality) across diverse domains. Although they have addressed NIDS, the scope of their survey is much broader. On the other hand, in \cite{luayTemporalAnalysisNetFlow2025}, the authors present an initiative to enrich traditional NetFlow and packet-based datasets with temporal attributes such as inter-arrival time and flow timing. The introduction of temporal feature-rich datasets provides a foundation for analyzing evolving network states rather than isolated observations.

In summary, previous surveys have provided comprehensive analyses of algorithms, datasets, and evaluation methodologies, yet none have systematically categorized NIDS according to their temporal processing mechanisms. The increasing interest in sequential learning and temporally enriched datasets underscores the relevance of the present work in establishing a conceptual foundation for temporal analysis to advance robust and proactive intrusion detection.

\section{Taxonomy} \label{sec:taxonomy}

This section introduces a novel temporal analysis framework and taxonomy developed in this study to systematically categorize time-aware NIDS approaches.
To the best of our knowledge, this is the first systematic classification explicitly focused on the temporal methodologies for NIDS. Furthermore, this section describes different dimensions, such as the detection mode and the classification type.

\subsection{Temporal Dimension}

An NIDS receives and analyzes network traffic to detect intrusions. Among the most common types of data it processes are network packets and traffic flows. The former can be captured in real time and includes information such as source and destination IP addresses and ports, protocol, and the packet payload. On the other hand, a traffic flow is a set of network packets that share the same attributes and occur within a specific time interval. These shared attributes are usually the source and destination IP addresses~\cite{liu2021fast}, or the 5-tuple \textit{(source IP, destination IP, source port, destination port, protocol)}~\cite{han_network_2023,li_dfaid_2022}. Thus, a traffic flow represents a summary of certain properties of network connections, such as duration, amount of data, or the number of packets transferred, within a given time interval. Moreover, a flow can be either uni or bidirectional.

A time series in the context of NIDS refers to a sequence of chronologically ordered network traffic observations, where each observation is associated with a specific timestamp. These observations may represent different levels of granularity, such as aggregated network metrics like the number of flows or volume of bytes transferred, or specific characteristics of individual flows. Temporal dependencies are assumed to exist, meaning that the current state of the network can be influenced by its past behavior. The observation interval may be regular (e.g., 1-second windows) or irregular, depending on the model and type of events being analyzed. Additionally, a \textit{time window} is a segment of such a series that groups all observations occurring within a defined time interval. In the case of NIDS, these windows can be used to extract global features from the observed traffic, build snapshots of the network state, or generate sequences for temporal modeling.

Therefore, to analyze the influence of time in an NIDS, it is necessary to rely on a time series of data. Because NIDS can process different types of data, various strategies can be employed to handle temporal information, which are detailed below and summarized in Table~\ref{tab:temporal-taxonomy}.

\begin{figure}
    \centering

    \begin{subfigure}[t]{0.47\textwidth}
        \centering
        \includegraphics[width=\linewidth]{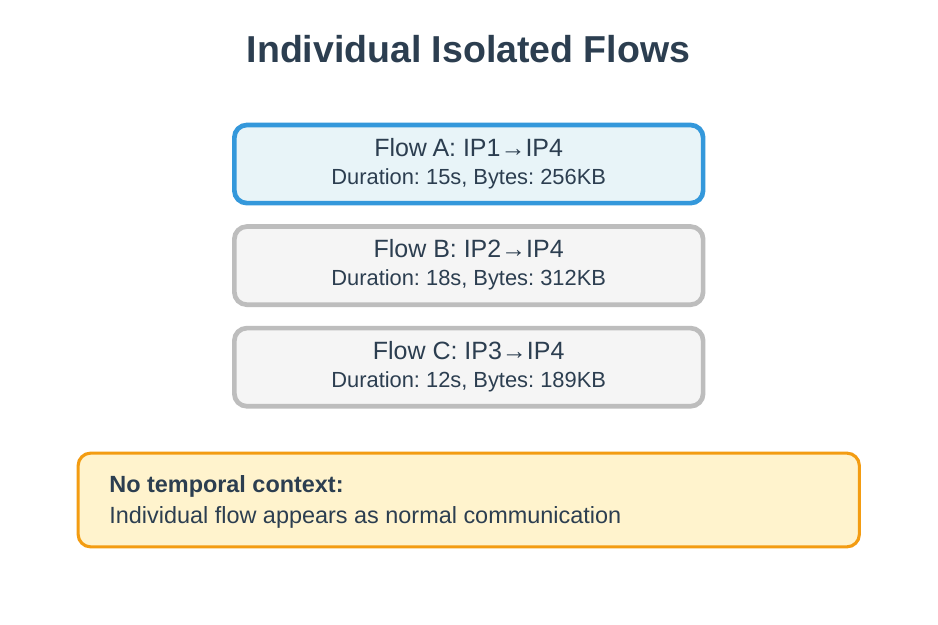}
        \caption{S.1 Static per-flow analysis. Each flow is analyzed independently; in this case, the first one analyzed (A) appears to be a normal communication.}
        \label{fig:S1}
    \end{subfigure}
    \hfill
    \begin{subfigure}[t]{0.47\textwidth}
        \centering
        \includegraphics[width=\linewidth]{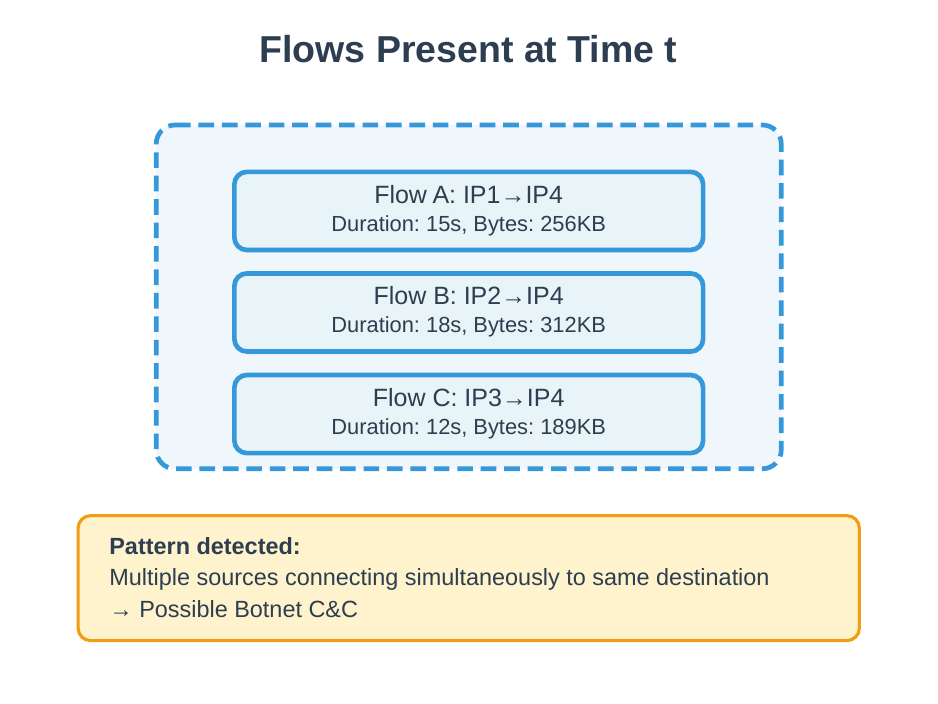}
        \caption{S.2 Static contextual snapshot analysis. All flows present at a given point in time (snapshot) are analyzed. Unlike \ref{fig:S1}, it is now observed that multiple sources are connecting to the same destination, which could indicate a case of C\&C.}
        \label{fig:S2}
    \end{subfigure}
   \caption{Taxonomy examples S.1 and S.2.}
    \label{fig:casos-estaticos}
\end{figure}

\paragraph{S.1 Static Per-Flow Analysis} Given the definition of traffic flow, one indirect way to approach temporality is to consider each flow individually, as illustrated in Figure~\ref{fig:S1}, where the analyzed flow (A) appears to represent normal communication. However, since flows are analyzed individually, no time series is formed. Even when analyzing multiple flows, the temporal sequence between them is not considered. An example of this approach is to use classifiers such as decision trees or logistic regression to classify each flow using all or a subset of its features (such as number of incoming/outgoing bytes or packets, or connection duration), as in~\cite{emanetEnsembleLearningBased2023}, where an ensemble of algorithms (logistic regression, decision trees, gradient boosting, among others) is used to vote on whether each flow is benign or malicious.

\paragraph{S.2 Static Contextual Snapshot Analysis} In this case, all flows or segments within a given time interval are grouped together for analysis, and global traffic features are extracted to classify them in isolation. The decision is based on information from other flows, i.e., not just the individual flow. Figure~\ref{fig:S2}, unlike Fig.~\ref{fig:S1}, shows multiple sources connecting simultaneously to the same destination, potentially indicating a C\&C case. For instance, by extracting aggregated features from the window, a GNN can be used to classify each flow or the entire snapshot by leveraging the network’s graph structure, as in~\cite{lo2021egraphsage}. Classifiers such as RF or SVM could also be used to classify flows based on tabular snapshot features, or anomaly detection methods may be applied. Temporality is implicitly present, as the snapshot is defined over a specific time interval, but no temporal sequence is analyzed to make decisions.

It is worth noting that some algorithms can belong to either S.1 or S.2, as DT o RF, depending on the structure of the input data: per-flow (individual analysis, S.1) or aggregated snapshots (contextual analysis, S.2).

\begin{figure}
    \centering

    \begin{subfigure}[t]{0.49\textwidth}
        \centering
        \includegraphics[width=\linewidth]{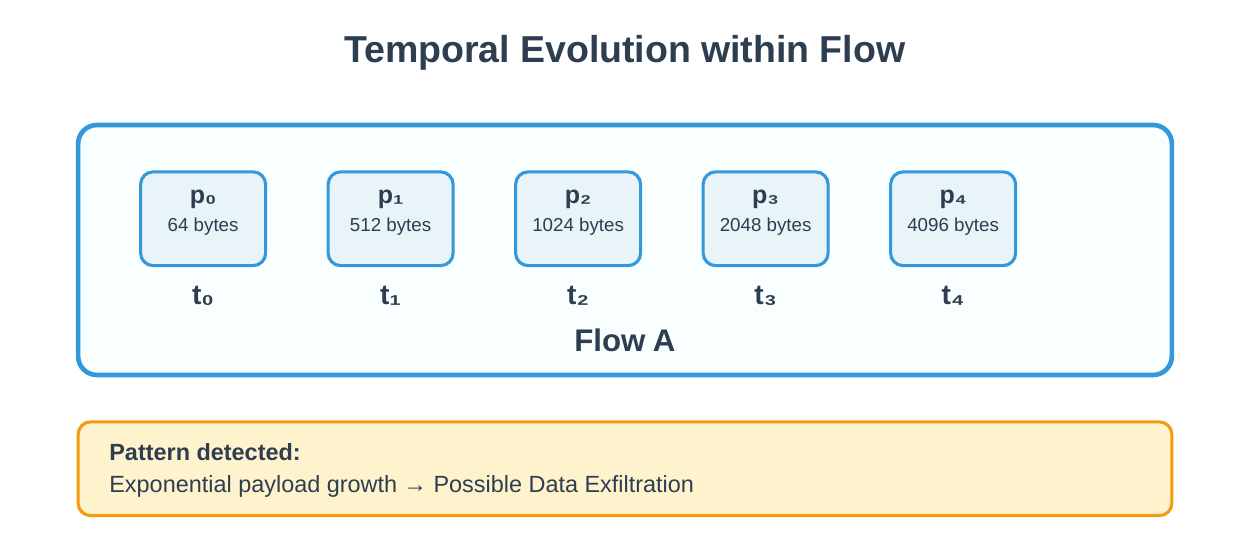}
        \caption{T.1 Intra-flow sequential analysis. It is observed that the flow packet size begins to increase over time, which would indicate a possible case of Exfiltration.}
        \label{fig:T1}
    \end{subfigure}
    \hfill
    \begin{subfigure}[t]{0.49\textwidth}
        \centering
        \includegraphics[width=\linewidth]{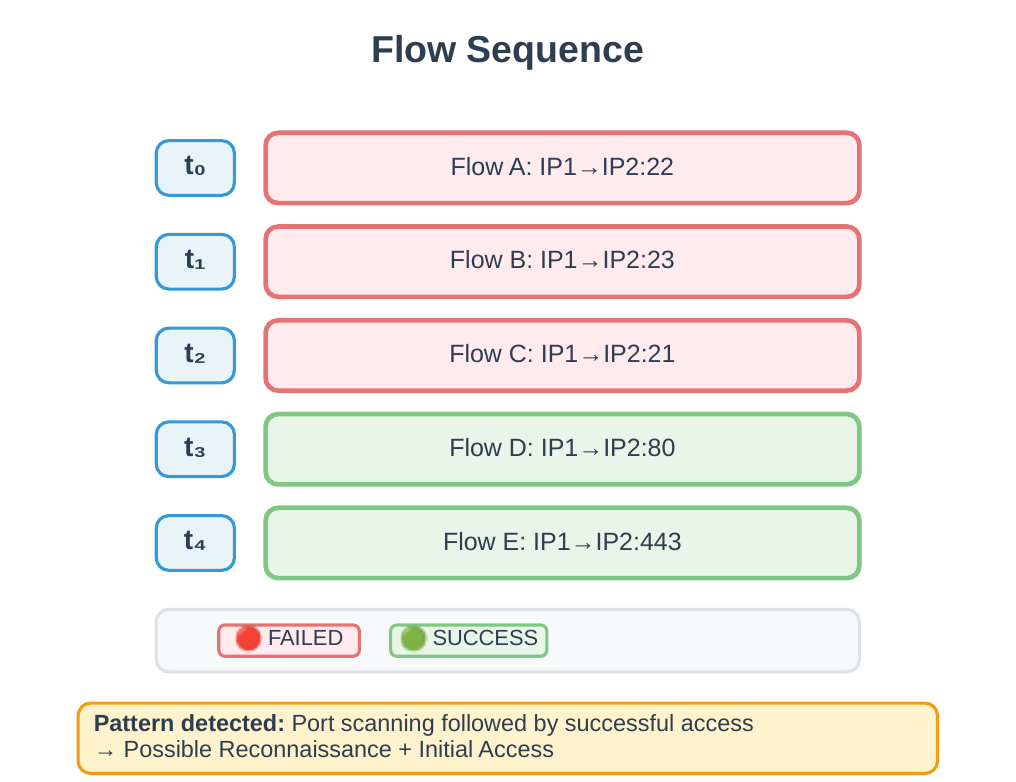}
        \caption{T.2 Inter-flow sequential analysis. A sequence of flows is observed from a single source (IP1), all directed to the same destination IP (IP2), but using different destination ports. This would indicate a case of Reconnaissance and Initial Access.}
        \label{fig:T2}
    \end{subfigure}

    \vspace{1em}

    \begin{subfigure}[t]{\textwidth}
        \centering
        \includegraphics[width=\linewidth]{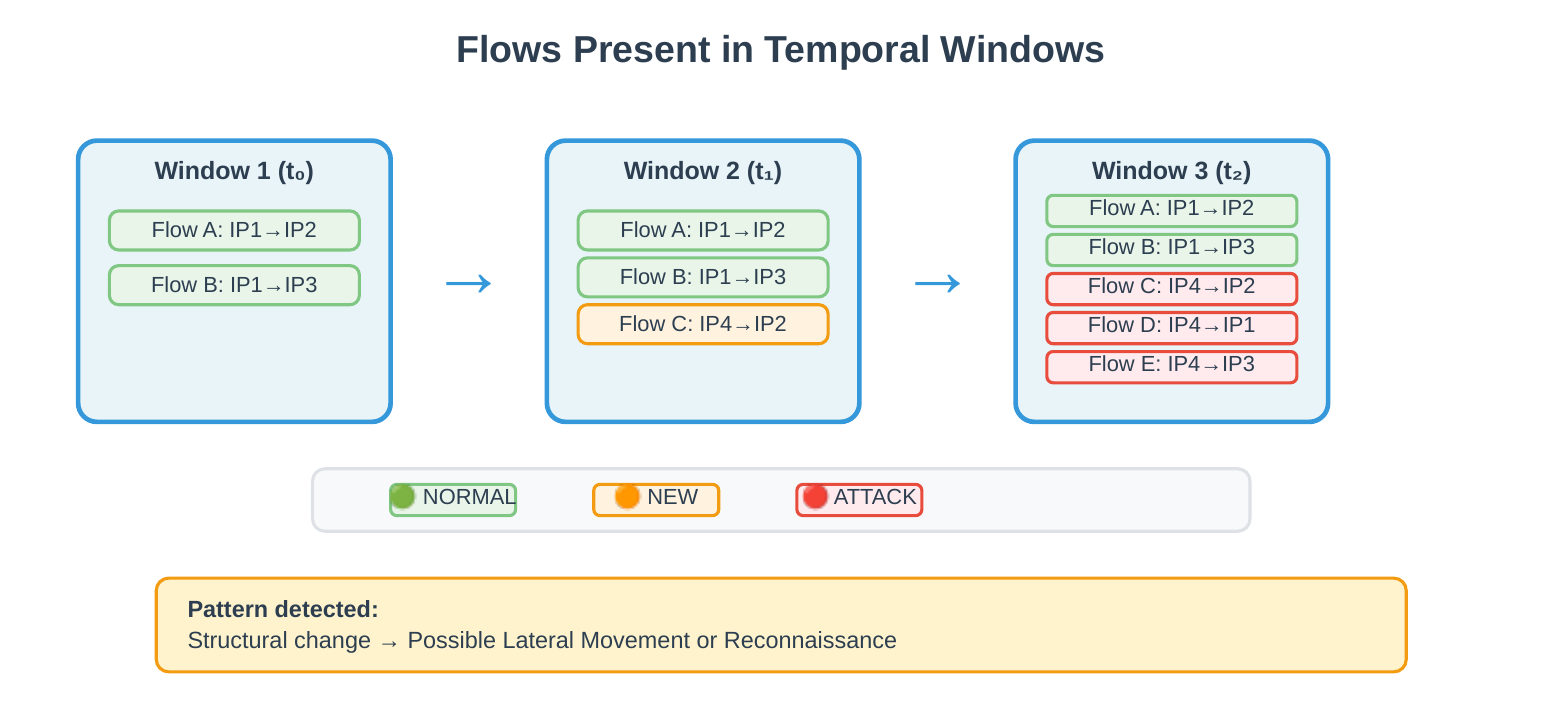}
        \caption{T.3 Temporal window-based sequential analysis. Looking at the succession of windows, we can see that a structural change occurs when a node and connections from that node to the rest are added, which would indicate a possible case of Lateral Movement or Reconnaissance.}
        \label{fig:T3}
    \end{subfigure}

    \caption{Case examples T.1, T.2, and T.3.}
    \label{fig:casos-temporales}
\end{figure}

\paragraph{T.1 Intra-Flow Sequential Analysis} This analyzes the temporal changes \textit{within the flow}, taking into account a sequence of features such as bytes or packets transferred for the same flow, as shown in Figure~\ref{fig:T1}. That is, it concerns intra-flow temporality. For example, in~\cite{chen_novel_2023}, the authors analyze the sequence of bytes transmitted for each flow organizing them into an array of 784 bytes. In this context, if we consider the flow from Figure~\ref{fig:T1}, where the connection consists of 5 packets at times $t_0$ to $t_4$ with sizes 64, 512, 1024, 2048, and 4096 bytes, respectively, the procedure to build the array is as follows: take all 64 bytes from the first packet, then all 512 bytes from the second, and the first 208 bytes from the third, thus obtaining 784 bytes in total. The remaining packets (2048 and 4096 bytes) are discarded, and if the total length of the flow is less than 784 bytes, it is padded with zeros at the end.

\paragraph{T.2 Inter-Flow Sequential Analysis} This approach models how flows behave in sequence, considering the temporal order in which they occur. It is essentially a temporally-aware version of the S.1 case, where temporality is present \textit{between flows}. For example, a flow sequence may be analyzed using a BiTCN network, as done in~\cite{cai_malicious_2024}. For illustration, Figure~\ref{fig:T2} shows five outgoing flows from IP1 to IP2 at different ports (22, 23, 21, 80, 443) at different times ($t_0$ to $t_4$), which may be interpreted as a port scan (Reconnaissance and Initial Access).

\paragraph{T.3 Temporal Window-Based Sequential Analysis} Instead of analyzing a static time window as in the S.2 case, this approach analyzes the temporal changes of the network using multiple time windows, as shown in Figure~\ref{fig:T3}. Various strategies exist, such as sliding windows (moving one step at a time along the time series), expanding windows (progressively increasing in size), or fixed windows (non-overlapping, independent segments). In this case, the time series is composed of a sequence of windows, making it a temporally-aware extension of S.2. For example, an RNN can be used to analyze the sequence of windows and detect structural changes, as done in~\cite{king2023euler}, where structural anomalies are detected, such as new or unexpected connections that violate normal communication patterns (e.g., a device accessing a resource without the usual prior authentication sequence).

\begin{table}
    \caption{Taxonomy of temporality.} \label{tab:temporal-taxonomy}
    \centering
    \scriptsize
    \begin{tabular}{@{}
    >{\centering\arraybackslash}p{1cm} 
      >{\raggedright\arraybackslash}p{3cm} 
      >{\raggedright\arraybackslash}p{3.5cm} 
      >{\centering\arraybackslash}p{2cm} 
      @{}}
    \toprule    
    \textbf{Code} & \textbf{Category} & \textbf{Description} & \textbf{Temporality} \\
    \midrule
    S.1 & Static Per-Flow Analysis & Individual flow analysis, no temporal context. & None \\
    S.2 & Static Contextual Snapshot Analysis & Multiple flows in time window, no sequences. & Implicit \\
    T.1 & Intra-Flow Sequential Analysis & Changes within single flow. & Local \\
    T.2 & Inter-Flow Sequential Analysis & Temporal sequences between multiple flows. & Global \\
    T.3 & Temporal Window-Based Sequential Analysis & Changes across multiple time windows. & Sequential \\
    \bottomrule
    \end{tabular}
\end{table}

\subsection{Detection Mode}

NIDSs commonly adopt two main approaches to identify intrusions: real-time detection and offline analysis. Their main characteristics are described below.

\paragraph{Real-Time} The focus is on identifying potential threats as quickly as possible in order to minimize potential damage. As indicated in~\cite{niknamiEnhancedMetaIDSAdaptive2025}, the detection time of a security incident depends on the effectiveness of the monitoring systems, as well as the configuration and capabilities of the NIDS. Furthermore,~\cite{jinSwiftIDSRealtimeIntrusion2020} points out that there are two challenges in this research: real-time IDS should be flexible and work on most commercial computers, and researchers should ensure that the system has excellent detection performance while achieving real-time capability. It is also worth noting that the increasing prevalence of encrypted network traffic often limits deep packet inspection, making efficient metadata-based detection a growing challenge for real-time systems.
 
\paragraph{Offline} Some slow, distributed, or stealthy attacks are more easily detectable through offline analysis. This approach involves analyzing stored network data or historical logs rather than processing packets in real-time. It is particularly useful for identifying persistent threats such as lateral movement or data exfiltration, which may evade immediate detection. Offline analysis is also commonly used in forensic investigations, where the goal is to reconstruct past events, improve existing detection rules, or validate hypotheses about attacker behavior. Moreover, it is often employed when computational resources are limited and performing full real-time analysis is not feasible.

\subsection{Classification Type}

An NIDS can adopt different detection strategies depending on its objectives:

\paragraph{Binary classification} It focuses on determining whether the observed network activity is benign or malicious. In this case, the system simply decides whether an intrusion has occurred, without distinguishing its specific nature. This approach is often preferred when the primary goal is fast detection with low computational cost and when false negatives (missed attacks) are more critical than understanding the exact attack type.

\paragraph{Multi-class classification} It aims to identify the specific type of intrusion once malicious activity has been detected. For example, the system might classify attacks into categories such as denial-of-service (DoS), port scanning, data exfiltration, or lateral movement. While this approach provides more detailed insights for incident response and forensic analysis, it generally requires more complex models, larger labeled datasets, and often results in higher computational overhead compared to binary classification.

In practice, some modern NIDS implementations combine both strategies, like \cite{niknamiEnhancedMetaIDSAdaptive2025}: they first perform a binary decision to filter potential threats and then apply multi-class classification only to the suspicious events, thereby optimizing both efficiency and accuracy.

\section{Literature Review} \label{sec:literature-review}

Taking into account the taxonomy proposed in Section \ref{sec:taxonomy}, a detailed analysis of the literature reviewed for this work is presented below. Table \ref{tab:papers} provides a summary of this analysis.

\begin{table}
    \caption{Literature review.} \label{tab:papers}
    \centering
    \scriptsize
    \begin{tabular}{@{}m{1cm}m{1cm}m{1cm}cm{6cm}cc@{}}
    \toprule
    \textbf{Group} & \textbf{Real-Time} & \textbf{Classification} & $\phantom{123}$ & \textbf{Method} & \textbf{Article} & \textbf{Year} \\
    \midrule
    \multirow{7}{*}{S.1} & \multirow{5}{*}{No} & \multirow{2}{*}{Binary} & & \multirow{2}{*}{Classic ML} & \cite{fengWeightedIntrusionDetection2021} & 2021 \\
     & & & & & \cite{emanetEnsembleLearningBased2023} & 2023 \\
     & & \multirow{3}{*}{Multiclass} & & ML & \cite{mbonaDetectingZeroDayIntrusion2022} & 2022 \\
     & & & & \multirow{2}{*}{ML and DL} & \cite{liuIntrusionDetectionImbalanced2021} & 2021 \\
     & & & & & \cite{aliDeepLearningVs2025} & 2025 \\
     \addlinespace
     & \multirow{2}{*}{Yes} & \multirow{2}{*}{Binary} & & Classic ML & \cite{sethNovelTimeEfficient2021} & 2021 \\
     & & & & DL & \cite{thirimanneDeepNeuralNetwork2022} & 2022 \\
     \addlinespace
     \addlinespace
    \multirow{4}{*}{S.2} & \multirow{3}{*}{No} & \multirow{3}{*}{\shortstack{Binary and\\multiclass}} & & \multirow{3}{*}{Static GNN} & \cite{lo2021egraphsage} & 2021 \\
    & & & & & \cite{xuApplyingSelfsupervisedLearning2024} & 2024 \\
    & & & & & \cite{dengEdgefeaturedMultihopAttention2025} & 2025 \\
    \addlinespace
    & Yes & Multiclass & & MPGNN & \cite{dingDivideConquerCoalesce2024} & 2024 \\
    \addlinespace
    \addlinespace
    \multirow{5}{*}{T.1} & \multirow{4}{*}{No} & Binary & & CNN + LSTM & \cite{shang_discovering_2021} & 2021 \\
     & & \multirow{3}{*}{Multiclass} & & 1D-CNN & \cite{Sadeghzadeh2021-Adversarial} & 2021 \\
     & & & & BiTCN & \cite{chen_novel_2023} & 2023 \\
     & & & & CNN + Transformer & \cite{wang2025imagtids} & 2025 \\
     \addlinespace
     & Yes & Binary & & CNN & \cite{Ghadermazi2025} & 2025 \\
     \addlinespace
     \addlinespace
     \multirow{5}{*}{T.2} & \multirow{3}{*}{No} & \multirow{2}{*}{Multiclass} & & BiTCN + attention & \cite{cai_malicious_2024} & 2024 \\
     & & & & 1D-CNN + LSTM teacher-student & \cite{wang_spatial-temporal_2024} & 2024 \\
     & & \multirow{1}{*}{\shortstack{Binary and\\multiclass}} & & CNN + self-attention & \cite{ren_canet_2023} & 2023 \\
     \addlinespace
      & \multirow{2}{*}{Yes} & Binary & & LSTM-based encoder-decoder & \cite{zavrakFlowbasedIntrusionDetection2023} & 2023 \\
      & & Multiclass & & Timestamp smoothing + DNN & \cite{han_network_2023} & 2023 \\
     \addlinespace
     \addlinespace
     \multirow{23}{*}{T.3} & \multirow{15}{*}{No} & \multirow{7}{*}{Binary} & & Statistical analysis + Chaos theory & \cite{fouladi2020ddos} & 2020 \\
     & & & & OCNN + LSTM & \cite{rajeshkannaUnifiedDeepLearning2021} & 2021 \\
     & & & & DCT-GAN & \cite{li2021dct} & 2021 \\
     & & & & Temporal transformer + encoder-decoder & \cite{wang_robust_2023} & 2023 \\
     & & & & GNN + GRU & \cite{king2023euler} & 2023 \\
     & & & & RNNs & \cite{syed_fog-cloud_2023} & 2023 \\
     & & & & GAN (LSTM, 1D-CNN, and TCN layers) & \cite{da_silva_ruffo_generative_2025} & 2025 \\
     & & \multirow{5}{*}{Multiclass} & & Convolutional that capture temporality & \cite{o_lopes_network_2023} & 2023 \\
     & & & & CNN + LSTM & \cite{shi2023multimodal} & 2023 \\
     & & & & DenseNet169 + SATNet & \cite{saikamEESNNHybridDeep2024} & 2024 \\
     & & & & SCNN + BiLSTM & \cite{bukhari2024secure} & 2024 \\
     & & & & AE + LSTM + CNN & \cite{susilo2025intelligent} & 2025 \\
     & & \multirow{4}{*}{\shortstack{Binary and\\multiclass}} & & CNN + LSTM & \cite{halbouniCNNLSTMHybridDeep2022} & 2022 \\
     & & & & CNN + BiLSTM & \cite{wanjau_discriminative_2023} & 2023 \\
     & & & & CNN + BiLSTM + DQN & \cite{wu_active_2024} & 2024 \\
     & & & & CapsNets + RNN & \cite{logeswari2025comprehensive} & 2025 \\
     \addlinespace
     & \multirow{7}{*}{Yes} & \multirow{5}{*}{Binary} & & CNN + LSTM & \cite{kimAIIDSApplicationDeep2020} & 2020 \\
     & & & & WL for directed graphs & \cite{liu2021fast} & 2021 \\
     & & & & MSCNN + LSTM & \cite{zhangModelIntrusionDetection2020} & 2020 \\
     & & & & CNN + LSTM + attention & \cite{psychogyiosDeepLearningIntrusion2024} & 2024\\
     & & \multirow{3}{*}{\shortstack{Binary and\\multiclass}} & & LightGBM + temporal statistic features & \cite{jinSwiftIDSRealtimeIntrusion2020} & 2020 \\
     & & & & GNN + GRU & \cite{duan2023application} & 2023 \\
     & & & & DL & \cite{niknamiEnhancedMetaIDSAdaptive2025} & 2025 \\
    \bottomrule
    \end{tabular}
\end{table}

\subsection{Static Approach}

Although various algorithms for detecting intrusions have been proposed in recent years, it is still common to find papers that address this problem from a \textit{static} view of the network, i.e., analyzing nodes or flows independently of time and the sequence between events. So these works are able to identify the so-called signatures of known attacks, based on prior knowledge. For example, in Haugerud \textit{et al.} (2021)~\cite{haugerudDynamicScalableParallel2021} parallelize an NIDS using several nodes with Snort sensors\footnote{\url{https://www.snort.org/}}. Also by observing the behavior, in Feng and Dou (2021)~\cite{fengWeightedIntrusionDetection2021}, Liu \textit{et al.} (2021)~\cite{liuIntrusionDetectionImbalanced2021}, Seth \textit{et al.} (2021)~\cite{sethNovelTimeEfficient2021}, Mbona and Eloff (2022)~\cite{mbonaDetectingZeroDayIntrusion2022}, Emanet \textit{et al.} (2023)~\cite{emanetEnsembleLearningBased2023}, and more recently, Ali \textit{et al.} (2025)~\cite{aliDeepLearningVs2025} detect anomalies or classify between types of flows. 

\cite{fengWeightedIntrusionDetection2021,emanetEnsembleLearningBased2023} use a voting strategy between different ML algorithms, such as KNN, MLP, SVM, and DT, among others, to evaluate each flow. On the other hand, \cite{sethNovelTimeEfficient2021} select features with a combination of RF and PCA, which reduces the time needed for real-time detection, using LightGBM as the best algorithm for classification. While \cite{liuIntrusionDetectionImbalanced2021,aliDeepLearningVs2025} implement traditional machine learning algorithms as well as various deep learning models, in \cite{mbonaDetectingZeroDayIntrusion2022} perform a particular feature selection (Benford’s law) applied to the numerical per-flow attributes, and train semi-supervised ML models on benign flow to detect zero-day attacks.

Thirimanne \textit{et al.} (2022)~\cite{thirimanneDeepNeuralNetwork2022} also emphasize real-time detection. They implement a DNN and propose a workflow that begins with real-time packet capture and preprocessing, and then evaluates them using an API where the DNN is applied to each flow. Thus, all of these cases correspond to category (S.1), which analyzes by flow in a static manner.

Following the taxonomy, the following works are characterized by performing a static contextual snapshot analysis (S.2). Lo \textit{et al.} (2021)~\cite{lo2021egraphsage}, Ding \textit{et al.} (2024)~\cite{dingDivideConquerCoalesce2024}, Xu \textit{et al.} (2024)~\cite{xuApplyingSelfsupervisedLearning2024}, and Deng and Hang (2025)~\cite{dengEdgefeaturedMultihopAttention2025}, which all represent data using graphs. In~\cite{lo2021egraphsage}, the authors propose E-GraphSAGE, an intrusion detection system for IoT networks based on graph neural networks (GNNs). To construct the input graph, nodes represent \textit{(IP, port)} pairs, while edges correspond to network flows, with their associated features derived from the flow attributes. The model uses an adapted version of GraphSAGE \cite{hamilton2017graphSAGE}, which, instead of generating embeddings for the nodes of the graph, produces embeddings for its edges.

Based on~\cite{lo2021egraphsage}, in~\cite{dengEdgefeaturedMultihopAttention2025} construct a graph whose nodes are  \textit{(IP, port)} pairs and whose edges are the requests between them, which they process using an adaptation of E-GraphSAGE. They also process the nodes using a single-layer multi-hop message passing, and apply attention mechanisms at both the node and edge levels, and an MLP for the final classification. To train the model, they define the edge features as the flow attributes, while the node features are constructed taking into account the number of times they act as source or destination in network connections, differentiated by traffic types. In this case, the snapshot covers the maximum possible time period of the dataset. Although the model does not receive the entire graph for learning, but rather mini-batches, these are formed randomly, so the temporality of the connections present in each mini-batch is not taken into account. Also based in \cite{lo2021egraphsage}, in \cite{dingDivideConquerCoalesce2024} propose Meta Parallel Graph Neural Network (MPGNN) to establish a scalable NIDS for large-scale IoT networks, which partition a graph into multiple subgraphs in a way that maximizes the performance and efficiency.

Meanwhile, in~\cite{xuApplyingSelfsupervisedLearning2024} to construct the input graph, nodes represent IP addresses and edges correspond to the connections between them. Edge features are TCP markers, protocol type, and number of bytes, while node features are given by a vector of 1s. The authors propose a GNN that extracts features from the edges using attention mechanisms (NEGAT) and implement a self-supervised graph representation learning method to identify different types of attacks (NEGSC). It should be noted that they only use a sample of the datasets, so the snapshot includes all existing connections in that sample.

\subsection{Temporal Approach}

Moving on to the analysis of temporal variation, there are articles in categories (T.1), (T.2), and (T.3). Starting with the first, which considers intra-flow variation, there are Sadeghzadeh \textit{et al.} (2021)~\cite{Sadeghzadeh2021-Adversarial}, Shang \textit{et al.} (2021)~\cite{shang_discovering_2021}, Chen \textit{et al.} (2023)~\cite{chen_novel_2023}, and more recently Ghadermazi \textit{et al.} (2025)~\cite{Ghadermazi2025}, and Wang \textit{et al.} (2025)~\cite{wang2025imagtids}. 

In \cite{Sadeghzadeh2021-Adversarial}, divide the input space of DL-based network traffic classification into three categories: Packet Classification, where the byte sequence of a packet is given to a classifier and each packet is labeled, Flow Content Classification, where the byte sequence of the first $n$ packets of a flow is fed to a classifier and the class of each flow is predicted, and Flow Time Series Classification, where the sequence of statistical features of the first $m$ packets of a flow, such as packets size and inter-arrival times between packets, are passed to a classifier, and each flow is classified. Then they challenge the robustness of DL-based network traffic classifiers aginst Adversarial Network Traffic (ANT). On the other hand, in \cite{shang_discovering_2021}, temporal features are extracted with an LSTM for each flow by analyzing the series of transferred packets. They focus only on the detection of C\&C channels and unknown APT attacks, using different classical ML classifiers. 

In \cite{chen_novel_2023}, a BiTCN is used to extract temporal features for each flow, analyzing the byte sequence directly and in reverse. Each byte in a traffic flow can be part of a series of messages or packets that are sent in a specific order. In \cite{Ghadermazi2025}, analyze header and payload data, considering temporal connections among packets. For that, the authors transform the sequential packets into two-dimensional images and process them with a CNN to detect malicious activities. Similarly, ImagTIDS~\cite{wang2025imagtids} also creates images, but it uses GADF, a trigonometry-based time series mathematical method, for encoding, and a hybrid CNN-Transformer architecture for analysis.

If the analysis focuses on temporal variation between flows (T.2), articles such as Han \textit{et al.} (2023)~\cite{han_network_2023}, Ren \textit{et al.} (2023)~\cite{ren_canet_2023}, Zavrak and Iskefiyeli (2023)~\cite{zavrakFlowbasedIntrusionDetection2023}, Cai \textit{et al.} (2024)~\cite{cai_malicious_2024}, and Wang \textit{et al.} (2024)~\cite{wang_spatial-temporal_2024} are found. For instance, \cite{han_network_2023} is proposed as an NIDS for real-time detection, where they extract and process the timestamp of each packet within a session to derive temporal features, and utilize features from both the header and the payload using DNNs. 

Another real-time detector is proposed in \cite{zavrakFlowbasedIntrusionDetection2023}, based on aggregating the features of flows present in 10-second intervals to form temporal windows from several of these intervals (between 3 and 10). Each of these windows is analyzed independently, i.e., they are not observed in succession. An anomaly detector is trained using an encoder-decoder, based on time windows extracted solely from normal traffic. Meanwhile, in \cite{ren_canet_2023}, they also process multiple flows grouped within a time window, from which spatio-temporal features are extracted by combining convolutional layers and self-attention mechanisms. They also use a loss function that addresses the problem of imbalance, known as EQLv2.

In \cite{cai_malicious_2024}, images in png format are constructed from the data as part of the preprocessing. These images are then processed by a BiTCN network that combines temporal convolutional layers with residual layers, taking into account both the original and reverse directions of traffic, and finally attention layers are used for classification. Also noteworthy is the replacement of the traditional negative log likelihood loss function with the cross-entropy loss function, which performs better when dealing with unbalanced classes. Each image is treated independently, i.e., the model does not observe a sequence of images, but temporal features are extracted from each image based on the sequence of flows in each one.

On the other hand, \cite{wang_spatial-temporal_2024} adopts a teacher-student learning approach, in which a complex model (the teacher) generates accurate detections, and through the \textit{Knowledge Distillation} technique, a simpler model (the student) is trained to replicate the knowledge learned by the former. Both models capture spatial and temporal features present in the flows grouped into 10-second snapshots, using a combination of 1D-CNN and LSTM layers. In addition, they use the focal loss function, which is especially useful for addressing class imbalance problems.

Continuing now with category (T.3) of the proposed taxonomy, articles that analyze the variation of time windows are discussed. For instance, Fouladi \textit{et al.} (2020)~\cite{fouladi2020ddos}, Jin \textit{et al.} (2020)~\cite{jinSwiftIDSRealtimeIntrusion2020}, Kim \textit{et al.} (2020)~\cite{kimAIIDSApplicationDeep2020}, Zhang \textit{et al.} (2020)~\cite{zhangModelIntrusionDetection2020}, Li \textit{et al.} (2021)~\cite{li2021dct}, Liu \textit{et al.} (2021)~\cite{liu2021fast}, Rajesh Kanna and Santhi (2021)~\cite{rajeshkannaUnifiedDeepLearning2021}, Halbouni \textit{et al.} (2022)~\cite{halbouniCNNLSTMHybridDeep2022}, Duan \textit{et al.} (2023)~\cite{duan2023application}, King and Huang (2023)~\cite{king2023euler}, Lopes \textit{et al.} (2023)~\cite{o_lopes_network_2023}, Wang \textit{et al.} (2023)~\cite{wang_robust_2023}, Wanjau \textit{et al.} (2023)~\cite{wanjau_discriminative_2023}, Shi \textit{et al}. (2023)~\cite{shi2023multimodal}, Syed \textit{et al.} (2023)~\cite{syed_fog-cloud_2023}, Bukhari \textit{et al.} (2024)~\cite{bukhari2024secure}, Psychogyios \textit{et al.} (2024)~\cite{psychogyiosDeepLearningIntrusion2024}, Saikam and Ch (2024)~\cite{saikamEESNNHybridDeep2024}, Wu \textit{et al.} (2024)~\cite{wu_active_2024}, and recently, da Silva Ruffo \textit{et al.} (2025)~\cite{da_silva_ruffo_generative_2025}, Logeswari \textit{et al.} (2025)~\cite{logeswari2025comprehensive}, Niknami \textit{et al.} (2025)~\cite{niknamiEnhancedMetaIDSAdaptive2025}, and Susilo \textit{et al.} (2025)~\cite{susilo2025intelligent}.

This category includes papers that explicitly define time windows, but also those that organize data into implicit time windows, given by the time interval covered by each batch. For example, in \cite{niknamiEnhancedMetaIDSAdaptive2025}, they emphasize real-time detection and propose a model that analyzes batches of flows at different stages, combining logistic regression and MLPs, refining its prediction with an incremental learning approach.

In \cite{kimAIIDSApplicationDeep2020,zhangModelIntrusionDetection2020,rajeshkannaUnifiedDeepLearning2021, halbouniCNNLSTMHybridDeep2022,shi2023multimodal,susilo2025intelligent} implement an NIDS based on a neural network that combines CNN and LSTM layers, either directly or through variants, thus extracting spatial and temporal features from network traffic. 
In \cite{kimAIIDSApplicationDeep2020}, there are two CNNs before the LSTMs, while in \cite{halbouniCNNLSTMHybridDeep2022}, the combination of CNN and LSTM layers is repeated three times, varying the number of neurons and filters. In \cite{shi2023multimodal}, the authors combine a two-branch CNN with an LSTM to extract the spatio-temporal feature information of network traffic, and in \cite{susilo2025intelligent} extract features with an autoencoder (AE), and then analyze them through an LSTM and CNN. In contrast, \cite{zhangModelIntrusionDetection2020} use a multiscale convolutional neural network (MSCNN) to combine with the LSTM, while \cite{rajeshkannaUnifiedDeepLearning2021} use a unified model of Optimized CNN (OCNN) and Hierarchical Multi-scale LSTM (HMLSTM), which allows it to learn more complex and detailed representations of temporal information. In turn, both \cite{kimAIIDSApplicationDeep2020} and \cite{zhangModelIntrusionDetection2020} raise the possibility of being used as real-time detectors, and even \cite{kimAIIDSApplicationDeep2020} help write and improve Snort rules for signature-based identification systems, based on the patterns they identify.

Similar to the previous ones, in \cite{logeswari2025comprehensive,syed_fog-cloud_2023,bukhari2024secure}, the authors also capture the spatio-temporal features, but using a combination of CapsNets and RNNs in \cite{logeswari2025comprehensive}, a feature selection followed by a SimpleRNN or a BiLSTM in \cite{syed_fog-cloud_2023}, and combining stacked CNNs (SCNN) with BiLSTM in \cite{bukhari2024secure}. Meanwhile, in \cite{wu_active_2024}, the authors propose a framework based on active learning that uses a DQN to detect zero-day attacks. It consists of three parts: an NIDS classifier, a sample selection strategy, and a labeler. For the classifier, a CNN-BiLSTM network is implemented, extracting both spatial and temporal features. For sample selection, a DQN-based strategy that consists of two LSTM networks to capture temporal features is used. 

In \cite{o_lopes_network_2023}, the authors are motivated by a previous research that models network intrusion detection as a time series task. They apply recent temporal convolutional architectures, such as MiniRocket, OS-CNN, TST, and XCM, that have shown success in other domains. These models use convolutional operations to capture temporal relationships by extracting complex features and patterns from network data sequences, without relying on traditional recurrent networks. In contrast, in \cite{saikamEESNNHybridDeep2024}, spatial and temporal features are captured using the DenseNet169 and SAT-Net models, respectively. In this article, the authors focus on the class imbalance problem, so the NIDS they propose performs \textit{hybrid} sampling, combining undersampling techniques (DSSTE) to reduce the majority class with oversampling techniques (DCGAN) to increase the minority class. 

On the other hand, articles such as \cite{jinSwiftIDSRealtimeIntrusion2020,wang_robust_2023,wanjau_discriminative_2023,psychogyiosDeepLearningIntrusion2024,da_silva_ruffo_generative_2025,li2021dct,syed_fog-cloud_2023} use sliding windows in their analysis, while in \cite{fouladi2020ddos} they use fixed windows of one minute each, and propose a DDoS attack detection and defense scheme based on statistical analysis and chaos theory. In \cite{jinSwiftIDSRealtimeIntrusion2020}, they propose SwiftIDS, which is a real-time detector that combines the use of the LightGBM algorithm to reduce preprocessing time and a parallel detection mechanism that divides the data into successive 2-second time windows. The authors extract temporal attributes ``manually'', that is, computing statistical features by time windows. In contrast, \cite{wang_robust_2023} propose an unsupervised method that constructs temporal windows called \textit{contexts}, and uses transformers to extract temporal features. It works as an anomaly detector, since part of the input data is masked and the model learns to reconstruct what is missing: in case of normal traffic, the reconstruction resembles what has already been observed, but this is not the case when anomalies occur. In \cite{syed_fog-cloud_2023}, 

Combining the use of sliding windows and spatio-temporal feature extraction, there is \cite{wanjau_discriminative_2023}, where such extraction is performed using a 2D-CNN and a BiLSTM. On the other hand, in \cite{psychogyiosDeepLearningIntrusion2024} the authors point out that real-time detection has limitations, as it does not allow threats to be anticipated before they occur, since traditional IDS methods are mainly reactive, i.e., they respond to attacks once they have already occurred. As an alternative, they propose a \textit{proactive} approach, in which the model analyzes patterns in the data within a time window to anticipate the occurrence of possible attacks in the future. To do this, they perform feature selection using ANOVA, and the network architecture they propose combines CNN, LSTM, and attention. For example, they predict whether there will be malicious activity in the next $T$ steps, i.e., they seek to determine whether any of the next $T$ flows will be malicious. To do this, they sort the dataset according to start time and generate time windows of size $W$, which represent the model's input. These can be assembled considering the entire dataset or by groups defined according to the connections between two IP addresses, to monitor traffic between two entities.

The NIDS system proposed in \cite{li2021dct,da_silva_ruffo_generative_2025} are based on a GAN model. In \cite{li2021dct}, the authors propose Dilated Convolutional Transformer-based GAN (DCT-GAN), which consists of multiple generators and a single discriminator. Each generator learns information from a different-sized detection window by using a Dilated CNN to extract features. In \cite{da_silva_ruffo_generative_2025}, the GAN detects anomalies by analyzing network flows grouped into sliding time windows of $q$ seconds (with $q=8,16$), which are updated every second. To construct these windows, only the IP addresses and ports, both source and destination, of the flows present in the last $q$ seconds are considered. From these, the entropy of each of the variables is calculated, thus forming a 4-tuple that ``summarizes'' the flows of the respective window to be analyzed. The GAN generator learns to generate these 4-tuples from a random vector, while the discriminator is trained to distinguish between real and synthetic traffic, allowing anomalies to be detected when a window deviates from the behavior learned as normal, thus classifying the entire window as anomalous. They test different architectures for both the generator and the discriminator that allow temporal features to be captured (LSTM, 1D-CNN, or TCN). It is also noteworthy that the generation of synthetic data with the GAN helps to solve the class imbalance problem.

Finally, \cite{liu2021fast,duan2023application,king2023euler} are analyzed, which take into account both structural or topological and temporal information. In \cite{liu2021fast}, they propose a real-time detector that focuses on detecting DDoS attacks and is based on the analysis of network packets rather than aggregate flows. To do this, they use a temporal network graph model, in which they take 10-second time windows and construct unweighted directed graphs $G_t$ from the nodes $V_t$ and connections $E_t$ present in each interval, where the nodes are given by the IP addresses and the edges represent the network packets transmitted from the source IP to the destination IP. The similarity between $G_t$ and a dynamic set of normalized graphs representing normal network states is then calculated using an adaptation of the Weisfeiler-Lehman method \cite{huang2021WeisfeilerLehmanMethod} for directed graphs in order to detect anomalies.

In \cite{king2023euler}, the authors propose an NIDS called \textit{Euler}, which leverages the topological and temporal characteristics of the data to detect lateral movement. To do this, they separate the data into half-hour time windows, and for each of these snapshots, they construct a directed graph representing the state of the network at that moment, and then they process the snapshots in parallel using GraphSAGE to create an embedding for each one. To construct these graphs, the nodes are entities that represent users and computers within the network, whose features are given by a one-hot vector that indicates what type they are, while the edges represent authentication events between these entities, and their feature is a weight that reflects the normalized frequency of authentications between those nodes in that time interval. After obtaining the topological embedding of each snapshot, they are processed in sequential order using a GRU. The model is trained with normal traffic data until the first anomalous interaction occurs, then tested on the rest of the dataset. In addition, they mask 5\% of the edges of each snapshot to perform inductive validation during training. 

On the other hand, \cite{duan2023application} propose a real-time NIDS based on line graph construction. Starting from an original graph $G$, where nodes represent IP addresses and edges represent connections between them, they generate its corresponding line graph $L(G)$. In this transformation, the nodes of $L(G)$ correspond to the edges (i.e., flows) of $G$, and an edge exists between two nodes in $L(G)$ if the corresponding edges in $G$ share a common node (i.e., an IP address). Using sliding windows, they build sequences of these line graphs and process them with a combination of GCNII-based GNNs for spatial feature extraction and GCNII-GRU layers to capture the temporal dynamics of network traffic. Similar to \cite{wang_spatial-temporal_2024}, they apply focal loss to address class imbalance.

\section{Discussion} \label{sec:discussion}

As described in the previous section, there are several ways to address temporality, each with different advantages and disadvantages.

\subsection{Comparative Complexity}

Static per-flow analysis (S.1) treats each network flow independently, typically using traditional classifiers. This approach generally has low to medium computational complexity. Models like RF are relatively light and can be parallelized across flows, making the method scalable for high traffic volumes \cite{masarat2016modified}. These models are often simpler (fewer parameters) and faster to train than deep learning models, thus exhibiting low computational overhead \cite{aliDeepLearningVs2025}. However, S.1 has a limited modeling capacity. Because it uses only static features of each flow, this method fails to capture temporal relationships between events or flows. Many cyber-attacks unfold over time with sequential patterns, and traditional static feature extraction methods cannot capture these temporal dependencies, often leading to inadequate detection of complex attacks \cite{zhangReviewDeepLearning2025}. 

On the other hand, static contextual snapshot analysis (S.2) considers a broader context to improve detection, often employing deep learning models such as GNNs to analyze these snapshots. This approach incurs medium to high computational complexity, as the use of CNN/GNN means significantly more computations (e.g. matrix multiplications in CNNs, iterative neighborhood aggregations in GNNs) compared to simple classifiers. Deep learning models typically require significant computational resources \cite{aliDeepLearningVs2025}. Moreover, graph-based processing adds overhead, as constructing a graph from network data and running GNN layers can be expensive, especially as graph size grows \cite{zhong2024survey}. Overall, S.2 enables richer contextual modeling, capturing spatial relationships and communication patterns within a time window, at the cost of higher complexity. 

Intra-flow sequential analysis (T.1) focuses on the sequence of events within a single flow. It typically leverages sequence models to model each flow’s timeline. This approach has high computational complexity, as these models process data step by step, for instance each packet in the sequence is handled in order \cite{shang_discovering_2021}. This inherent sequential nature means that, in general, the computation cannot be fully parallelized in time \cite{chen_novel_2023}. Therefore, analyzing long sequences is time-consuming. 

In contrast, inter-flow sequential analysis (T.2) extends the notion of temporality beyond individual flows. Although it may employ similar sequential architectures such as LSTMs or CNNs, it typically incurs higher computational complexity because it models sequences of flows rather than packets \cite{aliDeepLearningVs2025}. This often demands additional mechanisms, such as attention \cite{cai_malicious_2024,ren_canet_2023} or encoder–decoder structures \cite{zavrakFlowbasedIntrusionDetection2023}, to effectively model flow interactions. 

Temporal window-based sequential analysis (T.3) is the most complex approach of those discussed. It combines aspects of S.2 and T.2, as the data is divided into snapshots over time, that are analyzed using deep learning models and higher-level sequential models to capture temporal patterns across windows \cite{kimAIIDSApplicationDeep2020,king2023euler}. This pipeline has very high computational complexity. On the one hand, each snapshot analysis is itself resource-intensive. On the other, the sequence model processes a series of snapshot outputs, introducing an additional sequential computation \cite{o_lopes_network_2023,niknamiEnhancedMetaIDSAdaptive2025}. Additionally, managing the temporal windows adds complexity in choosing window size, overlap, and alignment \cite{duan2023application,psychogyiosDeepLearningIntrusion2024}.  

\begin{figure}
    \centering
    \includegraphics[width=0.75\linewidth]{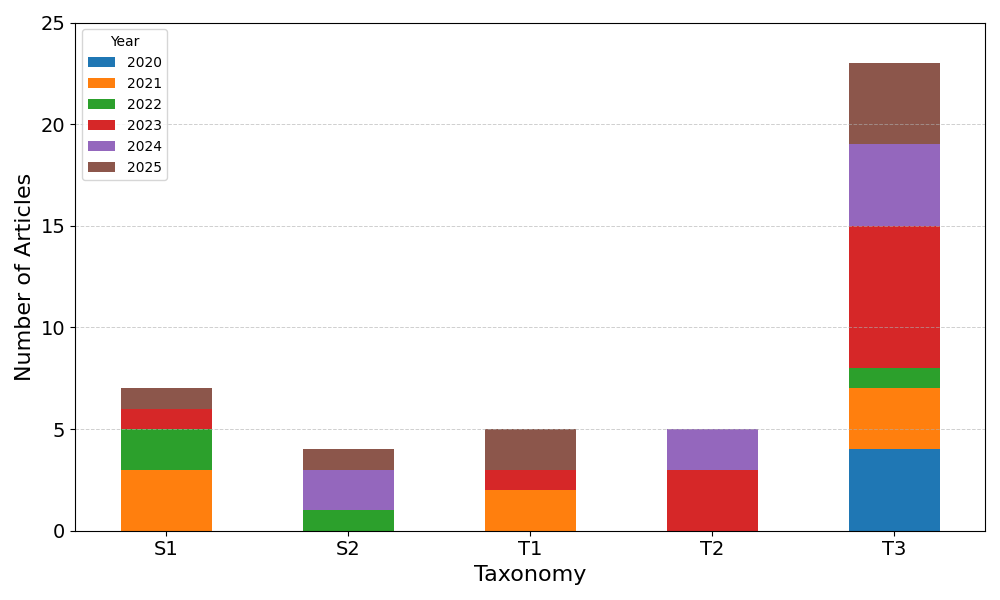}
    \caption{Number of articles that address temporality in different ways, broken down by year of publication.}
    \label{fig:articles-taxonomy}
\end{figure}

Beyond the challenges presented by sequential analysis, Figure \ref{fig:articles-taxonomy} shows that this is an area of interest in recent literature. It can also be seen that static per-flow analysis and by contextual snapshot continue to be addressed by authors, with the latter being the least explored approach. 

However, to detect threats early, it is still necessary to analyze behavior patterns over time, especially for prolonged attacks, i.e., those that are slow and low-level.

\subsection{Datasets}
\begin{figure}
    \centering
    \includegraphics[width=\textwidth]{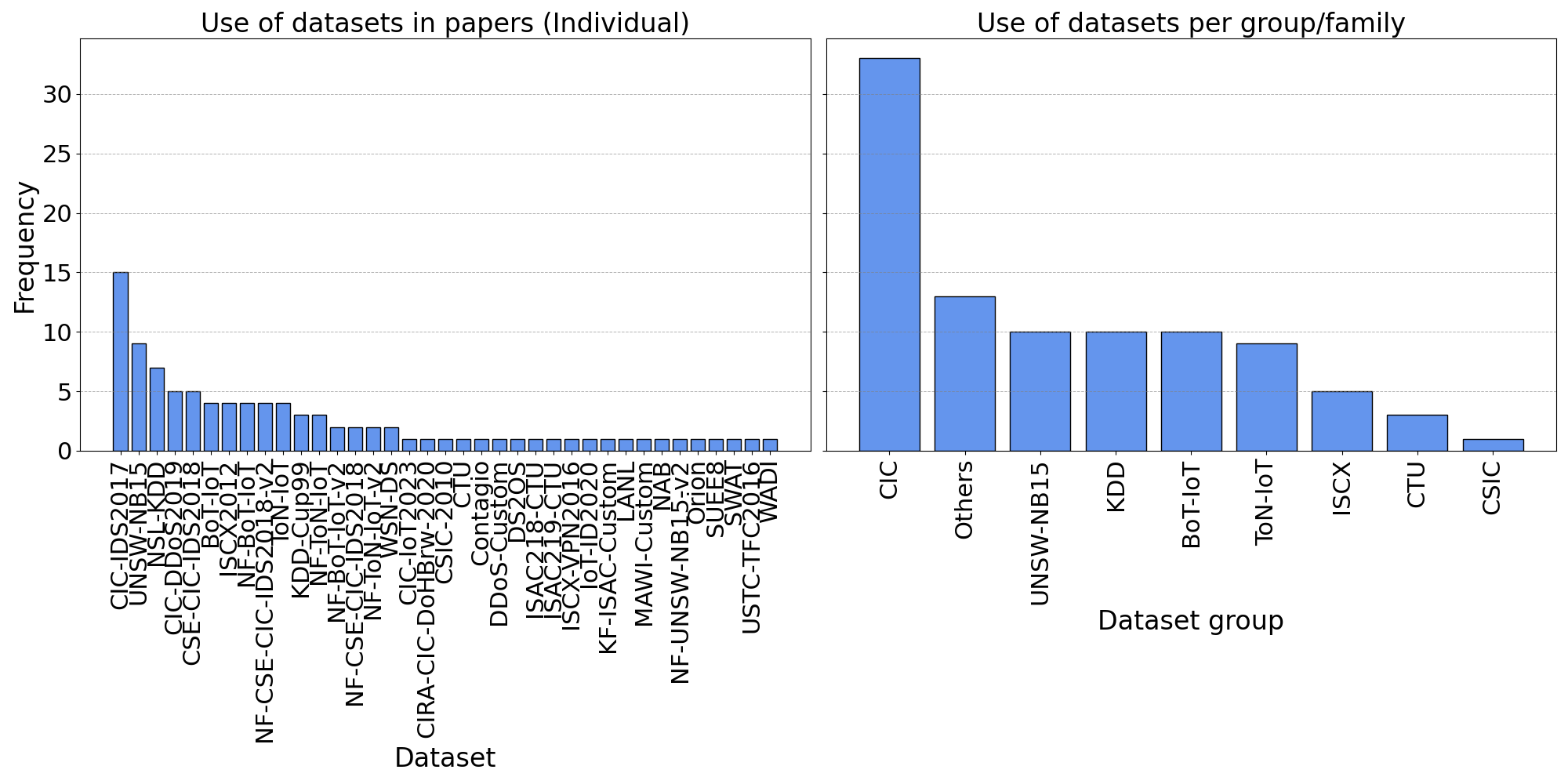}
    \caption{Datasets used in the literature analyzed. On the left, the frequency of each dataset is shown, while on the right, this is seen when grouped by family.}
    \label{fig:datasets}
\end{figure}

As Seth \textit{et al.} indicate in \cite{sethNovelTimeEfficient2021}, selecting the right dataset is essential for building an efficient machine learning model for intrusion detection. Although there are more than 80 datasets available for intrusion detection \cite{goldschmidt2025network}, the reviewed papers utilize only a small subset of them. As can be seen in Figure~\ref{fig:datasets} (left) CIC-IDS2017~\cite{sharafaldin2018-cic-ids}, UNSW-NB15~\cite{moustafa2015unsw}, and NSL-KDD~\cite{tavallaee2009-nsl-kdd} are the most commonly used datasets, and if grouped by data family (Fig.~\ref{fig:datasets} right), it can be seen that those generated by the Canadian Institute of Cybersecurity (CIC) are the ones mainly chosen.

\begin{figure}
    \centering
    \includegraphics[width=0.8\linewidth]{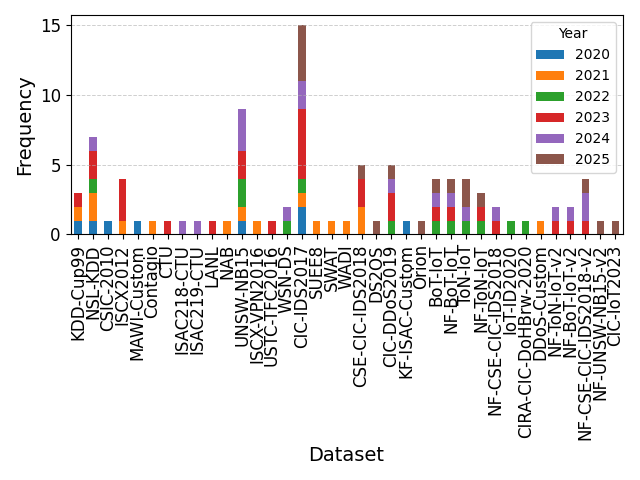}
    \caption{Number of articles that use the different datasets, broken down by year of publication. The order on the abscissa is ascending, starting with KDD-Cup99~\cite{kddcup99}, which is a 1999 dataset, and ending with CIC-IoT2023~\cite{Neto2023CICIoT2023} from 2023.} 
    \label{fig:datasets-article-year}
\end{figure}

While some of the aforementioned datasets are considered classic and serve to compare performance against other state-of-the-art NIDS, it is also true that evaluations on these datasets are not entirely meaningful, as they may be outdated and not reflect current traffic trends \cite{pintoReviewIntrusionDetection2025}. On the other hand, although UNSW-NB15~\cite{moustafa2015unsw} and CIC-IDS2017~\cite{sharafaldin2018-cic-ids} were released in 2015 and 2017, respectively, Figure~\ref{fig:datasets-article-year} shows that they remain widely used even in the latest studies (2024 and 2025). Nevertheless, several newer datasets have also been adopted, such as BoT-IoT (2019)~\cite{koroniotis2019botiot} and ToN-IoT (2020)~\cite{alsaedi2020toniot}, as well as their 2021 modified versions~\cite{sarhan2022modificaciones-datasets}, together with the modified versions of UNSW-NB15~\cite{moustafa2015unsw} and CSE-CIC-IDS2018~\cite{sharafaldin2018-cic-ids}, and more recently, the CIC-IoT2023 dataset (2023)~\cite{Neto2023CICIoT2023}.

\begin{figure}
    \centering
    \includegraphics[width=\linewidth]{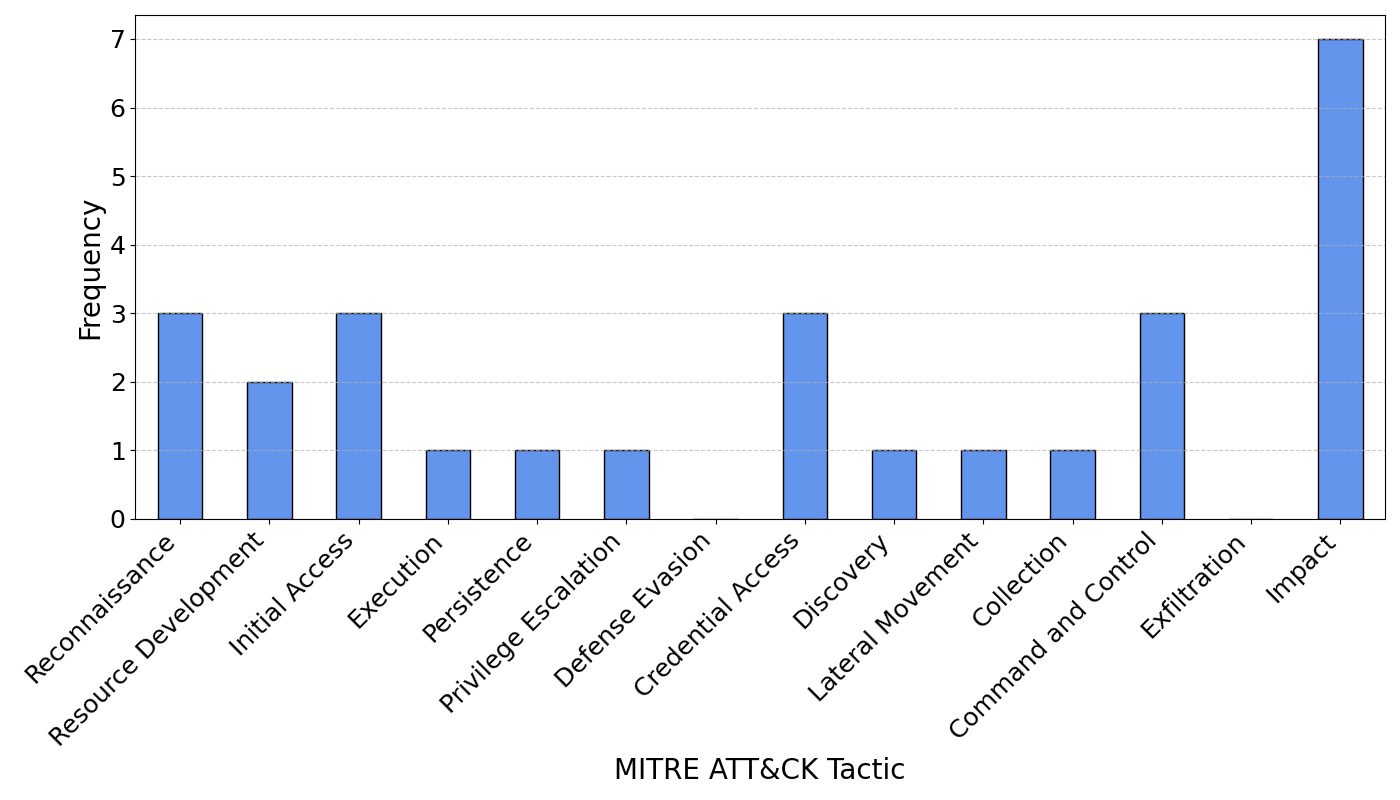}
    \caption{Frequency of MITRE ATT\&CK~\cite{mitreattack} tactics present among the CIC-IDS2017~\cite{sharafaldin2018-cic-ids}, NSL-KDD~\cite{tavallaee2009-nsl-kdd}, and UNSW-NB15~\cite{moustafa2015unsw} datasets.}
    \label{fig:mitre-top3datasets}
\end{figure}

Based on the literature review, CIC-IDS2017~\cite{sharafaldin2018-cic-ids}, NSL-KDD~\cite{tavallaee2009-nsl-kdd}, and UNSW-NB15~\cite{moustafa2015unsw} are the most commonly used datasets for intrusion detection and remain widely adopted in recent research. Accordingly, the specific types of attacks present in these datasets are analyzed below, using the MITRE ATT\&CK~\cite{mitreattack} framework as a reference for categorization. As can be seen in Figure~\ref{fig:mitre-top3datasets}, the tactic that is usually most represented is the last one, called \textit{Impact}. As described in Section~\ref{sec:mitre}, at this stage, the adversary attempts to manipulate, disrupt, or destroy the victim's systems and data, which is the worst-case scenario because it involves detecting when the most damage has already been done. This leads to training models that are biased toward detecting attacks after they have occurred, rather than recognizing adversary activity at early stages to prevent major damage \cite{psychogyiosDeepLearningIntrusion2024,pennaCTIHALHumanAnnotatedDataset2025}.

\subsection{Metrics} \label{sec:metrics-analysis}

\begin{figure}
    \centering
    \includegraphics[width=\linewidth]{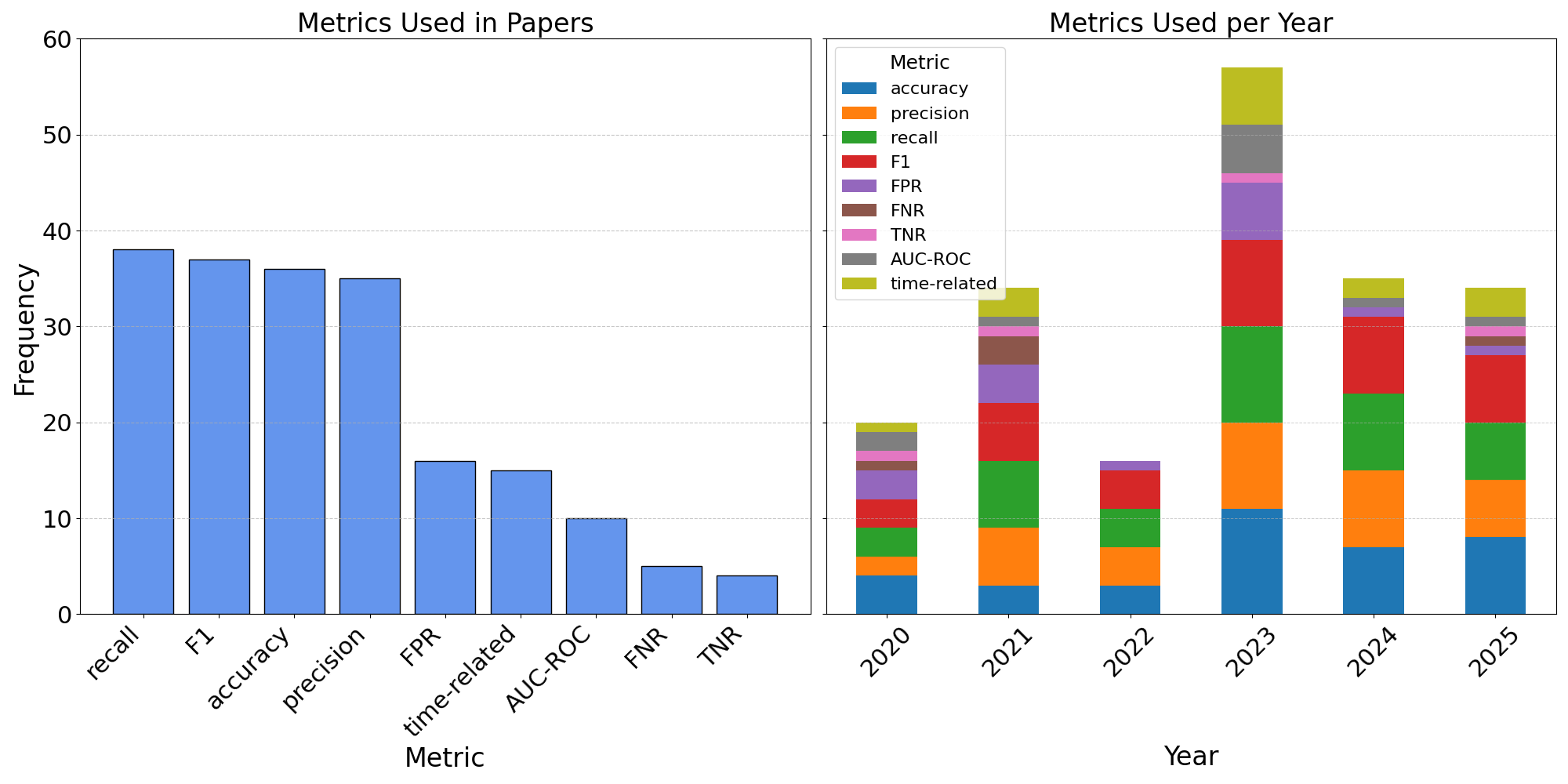}
    \caption{Frequency of metrics used in papers (left), broken down by year of publication (right).}
    \label{fig:metrics-papers-year}
\end{figure}

In order to understand how recent studies assess the performance of NIDS, the evaluation metrics reported in the reviewed papers are analyzed. As can be seen in Figure~\ref{fig:metrics-papers-year} (left), the most frequently reported metrics are conventional machine-learning performance indicators for classification, such as recall, F1, accuracy, and precision. Despite accuracy could produce inaccurate results in highly imbalanced datasets \cite{psychogyiosDeepLearningIntrusion2024}, as is common in NIDS, it continues to be used. Nevertheless, most articles complement it with additional metrics.

In contrast, metrics that capture false-positive and false-negative behavior, such as FPR, FNR, and TNR are rarely reported, even though these indicators reflect the reliability of a detection system. While traditional overall metrics can provide a general overview, they fail to account for consequences of misclassifications. That is, false positives result in  the use of additional resources, whereas false negatives can debilitate the entire system as they represent undetected intrusions \cite{sethNovelTimeEfficient2021}. On the other hand, time-related metrics, such as training time or inference time, are not often included, possibly because many of the reviewed studies do not perform real-time detection. Similarly, AUC-ROC, which is often considered a standard metric for evaluating classifiers under class imbalance conditions \cite{zavrakFlowbasedIntrusionDetection2023,wang_robust_2023}, is less commonly reported. 

Beyond conventional evaluation measures, the assessment of time-aware NIDS should incorporate metrics that explicitly capture temporal performance dynamics~\cite{tatbul2018precision}. For instance, metrics such as temporal precision and temporal recall, window consistency, or early detection rate could provide a richer characterization of NIDS behavior in dynamic, evolving network environments. Temporal precision and recall can quantify how effectively a model identifies attacks across consecutive time intervals rather than on isolated events; window consistency reflects the stability of predictions between adjacent temporal windows, indicating whether a detector maintains coherent behavior over time or exhibits erratic fluctuations; and the early detection rate provides a direct measure of how often attacks are identified before reaching late-stage tactics in the MITRE ATT\&CK lifecycle, emphasizing proactive over reactive detection. 

On the other hand, Figure~\ref{fig:metrics-papers-year} (right) presents the yearly evolution of metric usage between 2020 and 2025. Although there is a gradual increase in the diversity of reported metrics over time, the dominance of accuracy, precision, recall, and F1-score persists even in the most recent studies. This suggests that the research community still prioritizes general classification metrics over temporally aware or cost-sensitive evaluation, where error types are weighted according to their operational impact.

\subsection{Relevance of the Temporal Taxonomy for Attack Detection}

To assess the practical relevance of the proposed temporal taxonomy, a conceptual analysis of selected techniques from the MITRE ATT\&CK~\cite{mitreattack} framework was conducted. Although the evaluation was not empirical, this exploratory exercise highlights how various adversarial techniques may manifest patterns that are detectable through temporal analysis, thus motivating further investigation. 

This analysis focused on techniques that exhibit a temporal footprint, whether through repetitive, sequential, or time-windowed network activity. Techniques lacking clear temporal characteristics were excluded to maintain focus on the taxonomy’s applicability. A summary of this section is presented in Tables \ref{tab:mitre-taxonomy1} and \ref{tab:mitre-taxonomy2}.

Within the \texttt{TA0043 - Reconnaissance}~\cite{mitreTA0043-reconnaissance} tactic, \textit{Active Scanning} (\texttt{T1595}) is a technique that could be detected by analyzing temporality, since the adversary probes the victim's infrastructure through network traffic and can generate anomalous traffic or repetitive patterns over time. This could be examined by observing sequences of flows (T.2), or time windows (T.3).

\textit{Obtain Capabilities} (\texttt{T1588}) and \textit{Stage Capabilities} (\texttt{T1608}) are techniques from \texttt{TA0042 - Resource Development}~\cite{mitreTA0042-resource-development} tactic. During the acquisition, adversaries can purchase, download, or steal capabilities, i.e., tools, software, credentials, or technical resources that they can use to carry out their malicious activities in any phase of the attack. This technique could be detected with (S.2) or (T.3), provided that the acquisition comes from downloading executables or tools that are uncommon over time; however, it could not be detected directly if the downloads are legitimate. On the other hand, during the development or acquisition of capabilities, adversaries can upload, install, or set them up. This technique could be detected using (S.1) or (S.2) if there is an anomalous flow to repositories or platforms where tools are stored, such as GitHub, or with (T.2) or (T.3) if the adversary's procedure involves several steps.

If unusual traffic patterns are observed during time windows, such as multiple connections to suspicious domains just before malicious connections, cases of \textit{Phishing} (\texttt{T1566}) could be detected with (S.2) or (T.3). If adversaries attempt to exploit a vulnerability in a host or system connected to the Internet to initially access the network, then a case of \textit{Exploit Public-Facing Application} (\texttt{T1190}) could be detected by analyzing the presence of anomalies in each flow (S.1), such as anomalies in URLs, large or unusual payloads, or internal flow patterns (T.1), or if repeated attempts are made over time (T.2 or T.3). Similarly, the use of \textit{External Remote Services} (\texttt{T1133}), which are access mechanisms that allow users to connect to internal network resources from external locations, could be detected if anomalous traffic to services such as VPN, RDP, etc. is observed (S.2), or if access occurs with different temporal patterns (T.2 or T.3). These are techniques from \texttt{TA0001 - Initial Access}~\cite{mitreTA0001-initial-access}.

If adversaries abuse cloud administration services to execute commands within virtual machines, this could be detected only if they do so from an unexpected network. Then, techniques such as \textit{Cloud Administration Command} (\texttt{T1651}) could be detected by looking at snapshots (S.2). Similarly, if the adversary uses a remote container management API, then \textit{Container Administration Command} (\texttt{T1609}) could be detected if there is traffic to endpoints such as Docker or the Kubernetes API at unexpected times. Another technique from \texttt{TA0002 - Execution}~\cite{mitreTA0002-execution} that can be detected by analyzing the time is the one involving \textit{Scheduled Task/Job} (\texttt{T1053}), provided that these tasks make network connections, such as file transfers. If the tasks generate sequential traffic, (T.2) could be used, or (T.3) if the executions are repeated periodically in multiple windows.

\texttt{TA0003 - Persistence}~\cite{mitreTA0003-persistence} tactic consists of techniques that attackers use to maintain access to systems. Some techniques such as \textit{BITS Jobs} (\texttt{T1197}) can be detected by performing a temporal analysis, as adversaries abuse BITS (Windows Background Intelligent Transfer Service) jobs to persistently execute code and perform various tasks in the background. If BITS is used for network transfers, it can be detected through different approaches: (T.1) by analyzing intra-flow patterns, (S.2) by examining static contextual snapshots to check for unusual volumes of BITS traffic within a window, or (T.3) by analyzing temporal window sequences to identify periodic downloads. On the other hand, \textit{Traffic Signaling} (\texttt{T1205}) is used by adversaries to hide open ports or other malicious functions, which may consist of sending a series of packets with certain characteristics before a port is opened, which the adversary can use for command and control. This could be detected with (T.1) by analyzing the packet sequences within the same flow, or with (T.2) if the strategy involves Port Knocking, which is a covert communication technique that allows ports to be opened in a firewall only after receiving a specific sequence of connection attempts to certain closed ports. If the attempts occur periodically, they could be detected with (T.3).

\texttt{TA0004 - Privilege Escalation}~ \cite{mitreTA0004-privilege-escalation} tactic consists of techniques that adversaries use to obtain high-level permissions on a system or network. In general, adversaries can access and explore a network without privileges, but to achieve their objectives they require elevated permissions. Common approaches involve exploiting weaknesses, misconfigurations, and vulnerabilities in the system. These techniques often overlap with persistence techniques, as operating system functions that enable an adversary's persistence can be executed in an elevated context.

\texttt{TA0006 - Credential Access}~\cite{mitreTA0006-credential-access} consists of techniques for stealing credentials, such as account names and passwords. Among those detectable from temporal analysis (T.2 or T.3), the use of \textit{Brute Force} (\texttt{T1110}) and \textit{Exploitation for Credential Access} (\texttt{T1212}) stand out. Brute force techniques are used when passwords are unknown or when password hashes are obtained, using a repetitive or iterative mechanism to systematically guess them. By generating multiple authentication attempts (successful or unsuccessful) that are visible on the network, it is possible to detect login attempts distributed over time from the same IP or group (T.2), as well as to identify prolonged or slow brute force attacks (T.3). On the other hand, attackers can exploit software vulnerabilities to try to collect credentials, which allows them to execute code controlled by the attacker. This could be detected depending on whether the exploitation generates anomalous traffic or if there are repeated attempts over time.

\par The tactic \texttt{TA0007 - Discovery}~\cite{mitreTA0007-discovery} consists of techniques that an adversary can use to obtain information about the system and the internal network. Discovery techniques in various dimensions, such as \textit{Cloud Infrastructure Discovery} (\texttt{T1580}), \textit{Cloud Service Discovery} (\texttt{T1526}),  \textit{Network Service Discovery} (\texttt{T1046}), \textit{Network Share Discovery} (\texttt{T1135}), and \textit{Remote System Discovery} (\texttt{T1018}), are detectable if temporality is taken into account, as they can be analyzed by frequency or sequence. The discovery of cloud infrastructure and services could be detected using the (S.2) approach, if cloud APIs are used unusually from internal hosts, i.e., if there is unusual access to multiple cloud services, or with (T.3) if there is a sustained pattern of queries or scanning of services over time. In the case of network service discovery, it is highly detectable because port or service scans are visible, so a sequential analysis with (T.2) or (T.3) could help detect it. On the other hand, both the discovery of shared network resources and remote systems could be detected with (S.2) if anomalous connections to multiple shared resources are observed in the first case, or if there are many brief connections in the second case, or if the behavior persists or scales over time (T.3).

\texttt{TA0008 - Lateral Movement}~\cite{mitreTA0008-lateral-movement} consists of techniques that adversaries use to access and control remote systems on a network, some of which are detectable if temporality is taken into account. For example, with the use of \textit{Remote Services} (\texttt{T1021}), multiple flows would be observed, such as SSH connections, which can be seen as a suspicious pattern in a contextual snapshot (S.2), or can be analyzed as a sequence of remote logins (T.2). On the other hand, \textit{Exploitation of Remote Services} (\texttt{T1210}) could be detected by observing changes in intra-flow patterns (T.1) or the appearance of new unusual flows between nodes (T.2). Other techniques such as \textit{Lateral Tool Transfer} (\texttt{T1570}) and \textit{Software Deployment Tools} (\texttt{T1072}) could be detectable if transfer sequences or software distribution patterns are analyzed (T.2 or T.3).

Within the tactic \texttt{TA0009 - Collection}~\cite{mitreTA0009-collection}, some techniques are observed, such as \textit{Adversary-in-the-Middle} (\texttt{T1557}), \textit{Data from Information Repositories} (\texttt{T1213}), \textit{Data from Network Shared Drive} (\texttt{T1039}), and \textit{Automated Collection} (\texttt{T1119}). In the first technique mentioned, adversaries attempt to position themselves between two or more networked devices, which could be detected since interfering with a flow can cause detectable intra-flow changes (T.1), as well as alterations in the traffic pattern or between windows that can be detected with (T.3). On the other hand, the collection of data from information repositories and shared units on the network could be detected since repeated or multiple accesses to repositories or shared resources can form a detectable pattern in a snapshot (S.2) or as a sequence of flows (T.2). Also, automated collection could be detectable by analyzing successive windows (T.3), as it represents repetitive behavior over time.

\texttt{TA0011 - Command and Control}~\cite{mitreTA0011-CyC} consists of techniques that adversaries can use to communicate with systems under their control within the victim's network. Unlike the previous tactics, this one presents many more techniques that could be detected by an NIDS that analyzes temporality. For example, adversaries can continuously access and communicate with victims by injecting malicious content into systems through online network traffic (\textit{Content Injection}, \texttt{T1659}), or they can obfuscate command and control traffic to make it harder to detect (\textit{Data Obfuscation}, \texttt{T1001}). These techniques can be detected, as the former produces content alterations or insertions that may be visible in the flow (T.1) or in a sequence of windows (T.3), while in the latter, changes in packet size or header patterns can be detected if analyzed at the sequence level (T.2). On the other hand, adversaries can use an encryption algorithm to hide command and control traffic (\textit{Encrypted Channel}, \texttt{T1573}), use backup channels as alternative means of communication (\textit{Fallback Channels}, \texttt{T1008}), or use \textit{Multi-Stage Channels} (\texttt{T1104}) for command and control, all of which could be detected if the flow has patterns that may reveal anomalous behavior (S.1 or T.1), if changes in communication channels are observed in the sequence to be analyzed (T.2 or T.3), or if a progressive escalation of C2 is observed in multiple windows (T.3), respectively. Other techniques such as \textit{Ingress Tool Transfer} (\texttt{T1105}), whereby adversaries can transfer tools or other files from an external system to a compromised environment, or the use of \textit{Remote Access Tools} (\texttt{T1219}), whereby an adversary uses legitimate remote access tools to establish an interactive command and control channel within a network, may be detectable as they generate large or frequent transfers, or repetitive or persistent traffic patterns (T.1, T.2, or T.3). On the other hand, \textit{Traffic Signaling} (\texttt{T1205}) used to hide open ports or other malicious functions involves the use of a magic value or sequence that must be sent to a system to trigger a special response, so it can be detected by analyzing packet sequences (T.1), while the use of an external \textit{Web Service} (\texttt{T1102}) as a means of relaying data to or from a compromised system could be detected if a repetitive pattern is observed between flows to web services (S.2 or T.2).

Using \texttt{TA0010 - Exfiltration}~\cite{mitreTA0010-exfiltration}, the adversary attempts to steal data. The technique provided by \textit{Data Transfer Size Limits} (\texttt{T1030}), whereby the adversary can exfiltrate data in fixed-size fragments or below certain thresholds rather than complete files, could be detected by analyzing the intra-flow sequence (T.1) or the inter-flow sequence (T.2) because there is unusual consistency in the size or rate of the packets. If the exfiltration were done through a C2 channel, it could be detected by analyzing intra-flow anomalies (T.1) or inter-flow behavior (T.2). On the other hand, \textit{Exfiltration Over Alternative Protocol}, (\texttt{T1048}), \textit{Exfiltration Over Web Service} (\texttt{T1567}), or \textit{Transfer Data to Cloud Account} (\texttt{T1537}), could be detected by comparing snapshots (S.2), or by identifying unusual patterns over time (T.2 or T.3). Finally, if \textit{Scheduled Transfer} (\texttt{T1029}) is performed, it could be detected by analyzing the temporal regularity between flows (T.2) or time windows (T.3).

Last but not least, the final tactic in the attack lifecycle is called \texttt{TA0040 - Impact}~\cite{mitreTA0040-impact}. Within this, \textit{Email Bombing} (\texttt{T1667}), where adversaries can flood target email addresses with an overwhelming volume of messages, and \textit{Endpoint Denial of Service} (\texttt{T1499}) or \textit{Network Denial of Service} (\texttt{T1498}), where adversaries degrade or block the availability of services for users, are detectable by analyzing flow sequences or windows (T.2 or T.3), since the former is observed as a large number of outgoing flows to SMTP servers in repetitive patterns, while DoS attacks show excessive traffic or anomalous patterns. On the other hand, the use of \textit{Data Encrypted for Impact} (\texttt{T1486}) could be detected as an abnormal increase in internal encrypted traffic or abrupt changes between flows (T.1 or T.2). Finally, if system resources are exploited to perform resource-intensive tasks (\textit{Resource Hijacking}, \texttt{T1496}), this could be detected if prolonged traffic to external destinations is generated or anomalous resource usage is performed (T.1 or T.3).

\begin{figure}
    \centering
    \includegraphics[width=\linewidth]{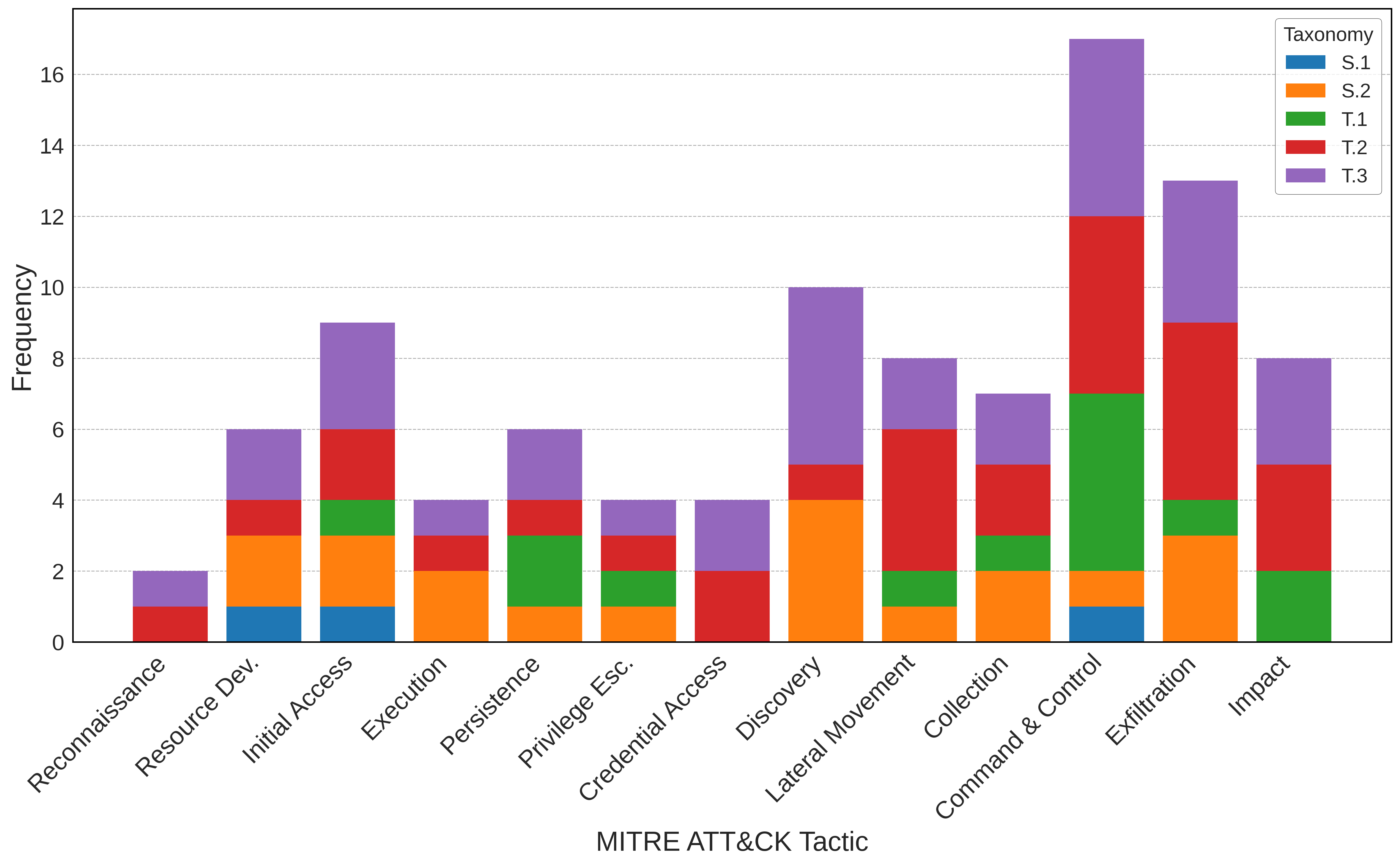}
    \caption{Number of attack techniques (MITRE ATT\&CK~\cite{mitreattack}) that could be detected by performing an analysis that follows one of the approaches described in the taxonomy proposed in this work.}
    \label{fig:taxonomies-by-tactic}
\end{figure}

Thus, it can be seen that several techniques have a temporal dimension that could be exploited to detect intrusions at different stages of the attack. As shown in Figure \ref{fig:taxonomies-by-tactic} and Tables \ref{tab:mitre-taxonomy1} and \ref{tab:mitre-taxonomy2}, approaches S.2, T.2, and T.3 are those that could be used to detect the greatest number of tactics. In particular, T.2 and T.3 seem to be the most versatile, as they could detect techniques from all the tactics analyzed. Although, as noted above, it is common to detect techniques from the Impact tactic, it would be interesting to be able to prevent them through early detection, making T.2 and T.3 the most recommended approaches for this task.

\begin{table}[H]
    \caption{Summary of the relationship between MITRE ATT\&CK~\cite{mitreattack} and the proposed taxonomy (Part 1).}
    \label{tab:mitre-taxonomy1}
    \centering
    \scriptsize
    \renewcommand{\arraystretch}{1.5}
    \begin{tabular}{@{}m{4cm} m{4.75cm} m{2.25cm} @{}}
    \toprule
    \textbf{Tactic} & \textbf{Technique} & \textbf{Taxonomy} \\
    \midrule
    TA0043 - Reconnaissance & T1595 - Active Scanning & T.2, T.3 \\
    \addlinespace
    
    \multirow{2}{*}{TA0042 - Resource Development} & T1588 - Obtain Capabilities & S.2, T.3 \\
    & T1608 - Stage Capabilities & S.1, S.2, T.2, T.3 \\
    \addlinespace
    
    \multirow{3}{*}{TA0001 - Initial Access} & T1566 - Phishing & S.2, T.3 \\
    & T1190 - Exploit Public-Facing Application & S.1, T.1, T.2, T.3 \\
    & T1133 - External Remote Services & S.2, T.2, T.3 \\
    \addlinespace
    
    \multirow{3}{*}{TA0002 - Execution} & T1651 - Cloud Administration Command & S.2 \\
    & T1609 - Container Administration Command & S.2 \\
    & T1053 - Scheduled Task/Job & T.2, T.3 \\
    \addlinespace
    
    \multirow{2}{*}{TA0003 - Persistence} & T1197 - BITS Jobs & S.2, T.1, T.3 \\
    & T1205 - Traffic Signaling & T.1, T.2, T.3 \\
    \addlinespace
    
    TA0004 - Privilege Escalation & General (overlapped with Persistence) & S.2, T.1, T.2, T.3 \\
    \addlinespace

    TA0005 - Defense Evasion & -- & -- \\
    \addlinespace
    
    \multirow{2}{*}{TA0006 - Credential Access} & T1110 - Brute Force & T.2, T.3 \\
    & T1212 - Exploitation for Credential Access & T.2, T.3 \\
    \addlinespace

    \multirow{5}{*}{TA0007 - Discovery} & T1580 - Cloud Infrastructure Discovery & S.2, T.3 \\
    & T1526 - Cloud Service Discovery & S.2, T.3 \\
    & T1046 - Network Service Discovery & T.2, T.3 \\
    & T1135 - Network Share Discovery & S.2, T.3 \\
    & T1018 - Remote System Discovery & S.2, T.3 \\
    
    \bottomrule
    \end{tabular}
\end{table}

\begin{table}[H]
    \caption{Summary of the relationship between MITRE ATT\&CK~\cite{mitreattack} and the proposed taxonomy (Part 2).}
    \label{tab:mitre-taxonomy2}
    \centering
    \scriptsize
    \renewcommand{\arraystretch}{1.5}
    \begin{tabular}{@{}m{4cm} m{4.75cm} m{2.25cm} @{}}
    \toprule
    \textbf{Tactic} & \textbf{Technique} & \textbf{Taxonomy} \\
    \midrule

    \multirow{4}{*}{TA0008 - Lateral Movement} & T1021 - Remote Services & S.2, T.2 \\
    & T1210 - Exploitation of Remote Services & T.1, T.2 \\
    & T1570 - Lateral Tool Transfer & T.2, T.3 \\
    & T1072 - Software Deployment Tools & T.2, T.3 \\
    \addlinespace
    
    \multirow{4}{*}{TA0009 - Collection} & T1557 - Adversary-in-the-Middle & T.1, T.3 \\
    & T1213 - Data from Information Repositories & S.2, T.2 \\
    & T1039 - Data from Network Shared Drive & S.2, T.2 \\
    & T1119 - Automated Collection & T.3 \\
    \addlinespace
    
    \multirow{9}{*}{TA0011 - Command and Control} & T1659 - Content Injection & T.1, T.3 \\
    & T1001 - Data Obfuscation & T.2 \\
    & T1573 - Encrypted Channel & S.1, T.1 \\
    & T1008 - Fallback Channels & T.2, T.3 \\
    & T1104 - Multi-Stage Channels & T.3 \\
    & T1105 - Ingress Tool Transfer & T.1, T.2, T.3 \\
    & T1219 - Remote Access Tools & T.1, T.2, T.3 \\
    & T1205 - Traffic Signaling & T.1 \\
    & T1102 - Web Service & S.2, T.2 \\
    \addlinespace
    
    \multirow{5}{*}{TA0010 - Exfiltration} & T1030 - Data Transfer Size Limits & T.1, T.2 \\
    & T1048 - Exfiltration Over Alternative Protocol & S.2, T.2, T.3 \\
    & T1567 - Exfiltration Over Web Service & S.2, T.2, T.3 \\
    & T1537 - Transfer Data to Cloud Account & S.2, T.2, T.3 \\
    & T1029 - Scheduled Transfer & T.2, T.3 \\
    \addlinespace
    
    \multirow{4}{*}{TA0040 - Impact} & T1667 - Email Bombing & T.2, T.3 \\
    & T1498/T1499 - Endpoint/Network DoS & T.2, T.3 \\
    & T1486 - Data Encrypted for Impact & T.1, T.2 \\
    & T1496 - Resource Hijacking & T.1, T.3 \\
    \bottomrule
    \end{tabular}
\end{table}

\section{Open Issues and Future Research Directions} \label{sec:open-issues}

Although the studies reviewed in this survey demonstrate clear progress in incorporating temporal reasoning into Network Intrusion Detection Systems (NIDS), several limitations persist across the literature. These shortcomings not only limit current performance but also outline promising avenues for future research. The most common limitations observed are summarized below, highlighting opportunities for advancing temporal NIDS toward practical, generalizable, and early detection.

\paragraph{Need for early detection and appropriate datasets} As noted in Section~\ref{sec:discussion}, researchers continue to rely heavily on classic datasets such as CIC-IDS2017~\cite{sharafaldin2018-cic-ids}, UNSW-NB15~\cite{moustafa2015unsw}, and NSL-KDD~\cite{tavallaee2009-nsl-kdd}, even though some of them were generated more than a decade ago, and there are many new ones \cite{goldschmidt2025network}. This can lead to an overestimation of model performance, as they do not reflect the characteristics and patterns of current network traffic, nor the evolution of attackers' tactics.

On the other hand, the distribution of adversarial labeled tactics present in the datasets is unbalanced, as most of the tactics included belong to the MITRE ATT\&CK \textit{Impact}~\cite{mitreTA0040-impact} tactic. This could be due to labeling issues \cite{goldschmidt2025network}. Nevertheless, if all MITRE phases are present in a dataset but not correctly identified, it represents an issue anyway. This means that many models are trained to detect malicious behavior in phases where the damage has already been done. In the early stages, most things seem normal, so there are many more False Positives \cite{maseerMetaanalysisSystematicReview2024,sunGNNIDSGraphNeural2024}, but it is important to recognize the sequence of events to detect attacks as soon as possible.

Future work should aim to construct temporally rich datasets covering entire attack lifecycles, with precise timestamps and clearly marked MITRE ATT\&CK phases. Furthermore, simulation platforms for persistent threats should be developed to generate realistic Advanced Persistent Threat (APT) scenarios and distributed attack campaigns that span extended time periods.

\paragraph{Little attention paid to class imbalance} One of the drawbacks inherent in this type of problem is the imbalance between classes, which is explicitly taken into account only by some of the papers analyzed. For example, in~\cite{kimAIIDSApplicationDeep2020,susilo2025intelligent,logeswari2025comprehensive,aliDeepLearningVs2025}, they oversample the minority class, while in~\cite{han_network_2023,bukhari2024secure} they subsample the majority class, in~\cite{sethNovelTimeEfficient2021,saikamEESNNHybridDeep2024,Sadeghzadeh2021-Adversarial,liuIntrusionDetectionImbalanced2021} they use a combination of both strategies, and in \cite{dingDivideConquerCoalesce2024} apply a stratified sampling (sampling the datasets based on their respective attack distributions). Even in works such as~\cite{zavrakFlowbasedIntrusionDetection2023,wang_spatial-temporal_2024,da_silva_ruffo_generative_2025,Ghadermazi2025}, they eliminate those attacks that are less frequent. However, these techniques may not be adequate, as they do not represent the underlying nature of a cyberattack~\cite{mudaNewOptimizedAdaptive2023}. 

On the other hand, other studies propose using different types of loss functions, such as focal loss~\cite{duan2023application,wang_spatial-temporal_2024,wang2025imagtids}, EQLv2 loss~\cite{ren_canet_2023}, or cross-entropy loss combined with attention mechanisms~\cite{cai_malicious_2024}, to deal with the imbalance problem. Meanwhile, in \cite{shi2023multimodal,mbonaDetectingZeroDayIntrusion2022}, they employ feature selection to improve the classification and identification of unbalanced network traffic data. Finally, in~\cite{da_silva_ruffo_generative_2025}, they address this by proposing a semi-supervised approach using a GAN,  while in works such as~\cite{zhangModelIntrusionDetection2020,xuApplyingSelfsupervisedLearning2024,dengEdgefeaturedMultihopAttention2025} they recognize the importance of the problem and propose it as future work. In contrast, the rest of the works either do not mention it or do not address it, which is striking given its extreme relevance to the problem.

A promising direction is to develop imbalance-aware learning techniques that preserve sequential dependencies, such as temporal data augmentation or generative models that create minority-class sequences while maintaining timing patterns.

\paragraph{Lack of generalization tests} A limitation found in the vast majority of the works analyzed is the lack of generalization tests for the proposed model. In this sense,~\cite{zhangModelIntrusionDetection2020,shang_discovering_2021,sethNovelTimeEfficient2021,thirimanneDeepNeuralNetwork2022,emanetEnsembleLearningBased2023,king2023euler,zavrakFlowbasedIntrusionDetection2023,psychogyiosDeepLearningIntrusion2024, niknamiEnhancedMetaIDSAdaptive2025,Sadeghzadeh2021-Adversarial,fouladi2020ddos, susilo2025intelligent,syed_fog-cloud_2023,aliDeepLearningVs2025} evaluate their models on a single dataset, whereas the remaining articles consider multiple ones, training each model independently on every dataset. The use of transfer learning mechanisms~\cite{weiss2016survey-TL,zhuang2020comprehensive-TL} is not even considered, except by \cite{logeswari2025comprehensive} that proposes it as future work. This means that the NIDS proposed in each case cannot be evaluated directly on a new network, but rather requires training the model for that data. Even in works such as~\cite{duan2023application,king2023euler,lo2021egraphsage} a minimal change in the network, such as the addition or removal of a node in the same previously used network, would require retraining with these modifications. The only partial exceptions are given by~\cite{da_silva_ruffo_generative_2025,shang_discovering_2021,Ghadermazi2025}, since in \cite{da_silva_ruffo_generative_2025}, for each dataset, they train with data from one day or segment of a day and validate and test with other days or segments, which may involve other types of attacks. In \cite{shang_discovering_2021}, they assemble the training and testing sets with different types of malicious traffic to evaluate the model's ability to detect unknown C\&C channels. Finally, in \cite{Ghadermazi2025} test the trained network intrusion detector in a different dataset, i.e., in a new target network environment.

In line with the above, even with the explicit possibility of performing generalization tests, it can be observed, for example, that in~\cite{o_lopes_network_2023} they avoid using the predefined train/test split, since the training set contains only six attack types, whereas the test set includes twelve. Instead, they unify both sets and build the training data by sampling from all available attacks. 
Also, \cite{wu_active_2024} address zero-day attack detection by simulating unseen classes, training the model on incomplete subsets of NSL-KDD and UNSW-NB15, and evaluating their ADQN framework's ability to detect these hidden categories in the testing phase. However, validation was performed by training and testing models separately on each dataset, without employing a cross-domain evaluation strategy to test universality across different network environments. On the other hand, some studies \cite{jinSwiftIDSRealtimeIntrusion2020,kimAIIDSApplicationDeep2020,zhangModelIntrusionDetection2020,rajeshkannaUnifiedDeepLearning2021,shang_discovering_2021,halbouniCNNLSTMHybridDeep2022, thirimanneDeepNeuralNetwork2022, han_network_2023, wang_robust_2023, wanjau_discriminative_2023, zavrakFlowbasedIntrusionDetection2023, cai_malicious_2024, saikamEESNNHybridDeep2024, xuApplyingSelfsupervisedLearning2024, wang_spatial-temporal_2024,wu_active_2024,wang2025imagtids,bukhari2024secure} provide no clear distinction between validation and testing, and hyperparameter tuning may have been performed on the test set, risking overfitting and undermining the evaluation of generalization. Even in \cite{liu2021fast,zavrakFlowbasedIntrusionDetection2023,shi2023multimodal}, they do not specify how they separate the data for training and testing. 

On the other hand, in \cite{sethNovelTimeEfficient2021,psychogyiosDeepLearningIntrusion2024,dengEdgefeaturedMultihopAttention2025,logeswari2025comprehensive} they use the \textit{k-fold cross validation} strategy, which consists of dividing the dataset into $k$ equal parts (folds) and performing $k$ model trainings, each time with a different combination of training and validation, while in \cite{halbouniCNNLSTMHybridDeep2022,ren_canet_2023,emanetEnsembleLearningBased2023} they adopt the \textit{stratified k-fold} version that maintains the proportion of classes in each fold. However, only~\cite{halbouniCNNLSTMHybridDeep2022,dengEdgefeaturedMultihopAttention2025} include testing on a separate portion of the data.

Future research should investigate transfer-learning strategies that reuse temporal representations learned from one domain and fine-tune them in another, as well as cross-domain benchmarks that explicitly test temporal generalization.

\paragraph{Need for more evaluation metrics} A further limitation observed across the reviewed literature involves the evaluation methodologies used to assess detection performance. As shown in Sec.~\ref{sec:metrics-analysis}, most studies rely almost exclusively on standard classification metrics such as accuracy, precision, recall, and F1-score. While those indicators provide a general overview of model performance, they overlook the reliability of intrusion detection system, in particular the impact of false positives and false negatives \cite{sethNovelTimeEfficient2021}. Similarly, few studies include AUC-ROC, even though it is a well-established measure for imbalanced classification problems \cite{zavrakFlowbasedIntrusionDetection2023}. Furthermore, the assessment of time-aware NIDS should incorporate metrics that explicitly capture temporal performance dynamics, such as temporal precision and temporal recall, window consistency, or early detection rate.

Said metrics should be taken into account in future research, in order to train more reliable and temporally aware NIDS. On the other hand, real-time NIDS should also report metrics like time-to-detection or time inference.

\paragraph{Limitations in the replicability of experiments} In works such as \cite{o_lopes_network_2023,ren_canet_2023,wanjau_discriminative_2023,wu_active_2024,bukhari2024secure}, the exact size or technique for dividing and constructing the sequences or time windows is not detailed, which makes it difficult to replicate their results.

Encouraging the creation of open-source repositories for temporal NIDS, similar to those maintained in other machine learning subfields, could increase the progress in this area.

\section{Conclusions} \label{sec:conclusions}

This paper presents a systematic review of recent literature on intrusion detection considering the temporal dimension. To this end, a taxonomy was proposed that provides a structured framework for analysis, allowing the different approaches to be classified and compared. Although intrusion detection considering temporality has been studied by numerous authors, there are still fundamental challenges that prevent truly effective and timely detection of attacks.

One of the main problems identified is the limited ability of current NIDS to detect threats in the early stages of the attack cycle. As evidenced in the analysis of the most widely used datasets (CIC-IDS2017, NSL-KDD, and UNSW-NB15), there is a greater representation of techniques associated with the MITRE ATT\&CK framework's \textit{Impact} tactic, while early phases are less represented. This disproportion favors the development of models trained to identify attacks when the damage is already irreversible, relegating many NIDS to the role of forensic analysis tools rather than proactive defense mechanisms. For example, detecting a ransomware attack when files have already been encrypted, or identifying an exfiltration once data has been extracted, constitutes a critical failure from a cybersecurity perspective.

That is why new lines of research are needed to address current limitations. In particular, there is a need to develop NIDS capable of anticipating attacks by detecting realistic temporal patterns associated with different stages of attacks, as well as persistent threats, considering different execution speeds and taking into account the dynamic context of the network over time. To this end, it is important to create new datasets that include complete sequences of adversarial behavior, from initial reconnaissance to final impact. Likewise, more rigorous and temporally oriented evaluation metrics are required to accurately reflect the real value of intrusion detection systems in dynamic, time-evolving scenarios. Taking steps in these directions means moving toward more effective solutions for early threat detection in increasingly complex and changing environments.

\section*{Acknowledge}
This work was supported by the National Scientific and Technical Research Council (CONICET, Argentina) through a doctoral fellowship awarded to T.P.

\section*{Declaration of generative AI and AI-assisted technologies in the manuscript preparation process}

During the preparation of this work the authors used generative AI tools, GPT-5, Claude Sonnet 4.5, and DeepL, to enhance writing in the paper, most notably for language refinement and translation. After using these tools, the authors reviewed and edited the content as needed and take full responsibility for the content of the publication.

\bibliographystyle{elsarticle-num}
\bibliography{references.bib}

\end{document}